\documentclass[aps,prc,reprint,showpacs,groupedaddress,onecolumn]{revtex4-1}

\usepackage{graphicx,color} 
\usepackage{dcolumn}  
\usepackage{bm}       
\usepackage{amsmath} 
\usepackage[dvipsnames]{xcolor}

\begin{document}

\newcommand {\nc} {\newcommand}

\newcommand{\vv}[1]{{$\bf {#1}$}}
\newcommand{\ul}[1]{\underline{#1}}
\newcommand{\vvm}[1]{{\bf {#1}}}
\def\btau{\mbox{\boldmath$\tau$}}

\nc {\IR} [1]{\textcolor{red}{#1}}
\nc {\IB} [1]{\textcolor{blue}{#1}}
\nc {\IP} [1]{\textcolor{magenta}{#1}}
\nc {\IM} [1]{\textcolor{Bittersweet}{#1}}
\nc {\IE} [1]{\textcolor{Plum}{#1}}

\nc{\ninej}[9]{\left\{\begin{array}{ccc} #1 & #2 & #3 \\ #4 & #5 & #6 \\ #7 & #8 & #9 \\ \end{array}\right\}}
\nc{\sixj}[6]{\left\{\begin{array}{ccc} #1 & #2 & #3 \\ #4 & #5 & #6 \\ \end{array}\right\}}
\nc{\threej}[6]{ \left( \begin{array}{ccc} #1 & #2 & #3 \\ #4 & #5 & #6 \\ \end{array} \right) }
\nc{\half}{\frac{1}{2}}
\nc{\numberthis}{\addtocounter{equation}{1}\tag{\theequation}}
\nc{\lla}{\left\langle}
\nc{\rra}{\right\rangle}
\nc{\lrme}{\left|\left|}
\nc{\rrme}{\right|\right|}

\title{\textit{Ab initio} Leading Order Effective Potentials for Elastic Nucleon-Nucleus
Scattering}

\author{M. Burrows$^{(a)}$}
\author{R. B. Baker$^{(a)}$}
\author{Ch. Elster$^{(a)}$}
\author{S.P. Weppner$^{(b)}$}
\author{K.D. Launey$^{(c)}$}
\author{P.~Maris$^{(d)}$}
\author{G.~Popa$^{(a)}$}

\affiliation{(a)Institute of Nuclear and Particle Physics,  and
Department of Physics and Astronomy,  Ohio University, Athens, OH 45701,
USA  \\
(b)  Natural Sciences, Eckerd College, St. Petersburg, FL 33711,
USA \\
(c) Department of Physics and Astronomy, Louisiana State University,
Baton Rouge, LA 70803, USA\\
(d) Department of Physics and Astronomy, Iowa State University, Ames, IA 50011, USA \\
}

\date{\today}

\begin{abstract}
\begin{description}
\item[Background] Calculating  microscopic effective interactions (optical potentials) for
elastic nucleon-nucleus scattering has already in the past led to a large body of work.
For first-order calculations a nucleon-nucleon (\textit{NN}) interaction and 
a one-body density of the nucleus were taken as input to rigorous calculations 
of microscopic full-folding calculations.

\item[Purpose] Based on the spectator expansion of the multiple scattering series we employ a chiral next-to-next-to-leading order (NNLO) nucleon-nucleon interaction on the same footing in the structure as well as in the reaction calculation to obtain an in leading-order consistent effective potential for nucleon-nucleus elastic scattering, which includes the spin of the struck target nucleon. 

\item[Methods] The first order effective folding potential is computed by first deriving a nonlocal scalar density as well as a spin-projected momentum distribution. Those are then integrated with the off-shell Wolfenstein amplitudes $A$, $C$, and $M$. The resulting nonlocal potential serves as input to a momentum-space Lippmann-Schwinger equation, whose solutions are summed to obtain the nucleon-nucleus scattering observables.

\item[Results] We calculate elastic scattering observables for $^4$He, $^6$He, $^8$He, $^{12}$C, and $^{16}$O in the energy regime between 100 and 200~MeV projectile kinetic energy, and compare to available data. We also explore the extension down to about 70~MeV, and study the effect of ignoring the spin of the struck nucleon in the nucleus.

\item[Conclusions] In our calculations we contrast elastic scattering off closed-shell and open-shell nuclei. We find that for closed-shell nuclei the approximation of ignoring the spin of the struck target nucleon is excellent. We only see effects of the spin of the struck target nucleon when considering $^6$He and $^8$He, which are nuclei with a $N/Z$ ratio larger than 1.

\end{description}
\end{abstract}


\maketitle

\section{Introduction and Motivation}
\label{intro}

Elastic scattering of protons or neutrons from stable nuclei has traditionally played an important role in
determining either the parameters of phenomenological optical models or testing accuracy and validity of
microscopic models thereof. The latter was explored in the 1990s in a large body of work on
microscopic optical potentials in which `high-precision' nucleon-nucleon ($NN$) interactions and the density
of the nucleus were taken as input to calculating the leading-order term in either a Kerman-McManus-Thaler (KMT)
or Watson expansion of the multiple scattering series (see,
e.g.,~\cite{Crespo:1992zz,Crespo:1990zzb,Elster:1996xh,Elster:1989en,Arellano:1990xu,Arellano:1990zz}).
This work concentrated on doubly magic nuclei like $^{40}$Ca and $^{208}$Pb, for which e.g.
mean field calculations provided the nuclear densities.

The  development of  nucleon-nucleon ($NN$) and three-nucleon ($3N$) interactions derived
from chiral effective field theory 
(see, e.g.,
~\cite{Epelbaum:2014sza,Epelbaum:2014efa,Reinert:2017usi,Epelbaum:2019kcf,Machleidt:2011zz,Entem:2017gor})
together with the
utilization of massively parallel computing resources (e.g.,
see~\cite{LangrDDLT19,LangrDDT18,SHAO20181,CPE:CPE3129,Jung:2013:EFO}), have placed {\it ab initio}
large-scale simulations at the frontier of nuclear structure and reaction explorations. Among other
successful many-body theories, the {\it ab initio} no-core shell-model (NCSM) approach (see, e.g.,
\cite{Navratil:2000ww,Roth:2007sv,BarrettNV13,Binder:2018pgl}), has over
the last decade taken center stage in the development of microscopic tools for studying the 
structure of atomic nuclei. The NCSM
concept combined with a symmetry-adapted (SA) basis in the {\it ab initio} SA-NCSM
\cite{LauneyDD16} has further expanded the reach to the structure of intermediate-mass
nuclei~\cite{Dytrych:2020vkl}.
Following these developments in nuclear structure theory, it is worthwhile to
again consider the rigorous calculations of leading-order effective folding nucleon-nucleus ($NA$)
potentials, since now the one-body densities required for the folding with the $NN$ scattering
amplitudes can be based on the same $NN$ interaction~\cite{Gennari:2017yez,Burrows:2018ggt}. 

However, a closer inspection of the theoretical ingredients of the leading order term of the multiple
scattering expansion underlying this latest work as well as 
the works from the 1990s reveals that they can not be considered fully {\it ab initio}. 
Those calculations ignore the spin of the struck target nucleon in the derivation of the leading order
effective interaction. Thus, in~\cite{Gennari:2017yez,Burrows:2018ggt} all terms of the $NN$ interaction are
included in the structure calculation but not in the reaction calculation. In order to construct an \textit{ab initio} effective $NA$ interaction, the underlying $NN$ interaction must be taken into account on equal
footing in all parts of the calculations. 

The idea that an effective interaction may depend on the spin of the struck target nucleon was pioneered
in~\cite{Cunningham:2011zz,Cunningham:2013lga} in the context of spin-spin terms in elastic scattering
from a target with nonzero spin, in which the authors use as a starting point for the nuclear structure a core
and valence nucleons. However, even when considering scattering of a proton or neutron from a spin-zero
nucleus, as done in this work, the spin of the struck target nucleon should be taken into account when
employing the $NN$ interaction on the same footing as in a modern structure calculation.  
Starting from a NCSM, we calculate the one-body density for the spin of the struck target nucleon and combine it with the corresponding terms
in the $NN$ amplitudes. The first goal of this manuscript is the presentation of the theoretical
formulation of taking into account the spin of the struck nucleon in the well-known formalism of the
leading order term in the multiple scattering expansion, the second is a study of its effects on elastic scattering observables for closed as well as open-shell nuclei.

In Sec.~\ref{formal} we briefly connect to the scattering formalism for the leading order term in the multiple scattering series as given in~\cite{Burrows:2018ggt} and introduce the spin-dependent one-body density together with its spin projections onto the scattering plane needed to include the full $NN$ interaction into the leading order term. In Sec.~\ref{results} we present elastic scattering observables for closed and open-shell nuclei based on a consistent {\it ab initio} calculation and compare to calculations in which the spin of the stuck nucleon is ignored. We conclude in Sec.~\ref{conclusions}.


\section{Theoretical Framework}
\label{formal}

The standard starting point for describing elastic scattering within a multiple scattering approach is the separation of the Lippmann-Schwinger (LS) equation for the transition amplitude
\begin{eqnarray}
\label{eq1}
	T=V+VG_0(E)T
\end{eqnarray}
into two parts, namely an integral equation for $T$,
\begin{eqnarray}
\label{eq2}
	T=U + U G_0(E)PT~,
\end{eqnarray}
where $U$ is the effective potential operator defined by a second integral equation
\begin{eqnarray}
\label{eq3}
	U=V + V G_0(E)QU~.
\end{eqnarray}
Here $P$ is a projection onto the ground state of the target, $P=\frac{\left| \Phi
\right\rangle \left\langle \Phi \right|}{\left\langle \Phi  | \Phi
\right\rangle}$, with $P+Q=1$ and $[G_0(E),P]=1$. The free propagator for the projectile and
target system is given by $G_0(E)=\left( E - h_0 - H_A + i\epsilon \right)^{-1}$ where $h_0$
is the kinetic energy of the projectile and $H_A$ is the Hamiltonian of the target defined by
$H_A \left| \Phi \right\rangle = E_A \left| \Phi \right\rangle$. The potential operator
$V=\sum_{i=1}^A v_{0i}$ consists of the $NN$ potential $v_{0i}$ acting between the projectile
denoted by ``0'' and the $i$th target nucleon.

Working in leading order of the spectator expansion means taking only the interaction between
the projectile and one of the target nucleons into account. Thus, in leading order the
effective interaction is given by $U=\sum_{i=1}^A \tau_{0i}$, where the operator $\tau_{0i}$ is 
given by
\begin{eqnarray}
\label{eq4}
	\tau_{0i} &=& v_{0i} + v_{0i}G_0(E)Q\tau_{0i} \cr
	&=& \hat{\tau}_{0i} - \hat{\tau}_{0i} G_0(E) P \tau_{0i}.
\end{eqnarray}
The quantity $\hat{\tau}_{0i}$ is the solution of a standard LS equation with the $NN$ potential 
as the driving term. For the effective interaction only $\widehat{U}=\sum_i^A \hat{\tau}_{0i}$
needs to be calculated, and $U$ is then obtained by solving Eq.~(\ref{eq4}) with $\widehat{U}$ as
the driving term. Explicitly the leading order effective interaction $\widehat{U}$, which is nonlocal and energy dependent, can be symbolically written as
\begin{align*}
\label{eq5}
	\widehat{U}(\bm{q},\bm{\mathcal{K}}_{NA},\epsilon) = \sum_{\alpha=n,p} \sum_{K_s} \int d^3{\mathcal{K}} \; 
	&\eta\left( \bm{q}, \bm{\mathcal{K}}, \bm{\mathcal{K}}_{NA} \right) \;
	\hat{\tau}^{K_s}_\alpha \left( \bm{q}, \frac{1}{2}\left( \frac{A+1}{A}\bm{\mathcal{K}}_{NA} - \bm{\mathcal{K}} \right); \varepsilon \right) \\
	&\rho^{K_s}_\alpha \left(\bm{\mathcal{K}} - \frac{A-1}{A} \frac{\bm{q}}{2}, \bm{\mathcal{K}} + \frac{A-1}{A} \frac{\bm{q}}{2} \right). \numberthis	
\end{align*}
The sum over $\alpha=n$ for neutrons and $p$ for protons, indicates that e.g. for a proton as projectile, the $pp$ amplitudes are integrated with the proton density and the $np$ amplitudes with the neutron density. The variable $\varepsilon$ represents the beam energy of the projectile minus the kinetic energy of the center-of-mass of the interacting particle and the binding energy of the struck particle. The index $K_s$ is either 0 for spin-independent terms or 1 for spin-dependent terms.
The momentum vectors in Eq.~(\ref{eq5}) are defined as
\begin{eqnarray}
\label{eq6}
	\bm{q} &=& \bm{p'} - \bm{p} \cr
	\bm{\mathcal{K}} &=& \frac{1}{2} \left(\bm{p'} + \bm{p}\right) \cr
	\bm{\mathcal{K}}_{NA} &=& \frac{A}{A+1}\left[\left(\bm{k'} + \bm{k}\right) +
\frac{1}{2} \left(\bm{p'} + \bm{p}\right) \right],
\end{eqnarray}
where $\bm{p}$ ($\bm{p'}$) is the momentum of the struck target nucleon before (and after) the 
collision, and $\bm{k}$ ($\bm{k'}$) the momentum of the projectile before
(and after) the collision. The momentum transfer $\bm{q}$ is invariant between frames, however the other vectors given
in Eq.~(\ref{eq6}) are frame dependent. More details about the different momentum vectors in each frame
are discussed in Appendix A of Ref.~\cite{Orazbayev:2013}. The terms in Eq.~(\ref{eq5}) are the M\o ller factor~\cite{CMoller} 
$\eta$, describing the frame transformation relating the zero-momentum $NN$ frame
to the zero-momentum $NA$ frame, the $NN$ amplitude between the projectile and the target nucleon,
$\hat{\tau}^{K_s}_{\alpha}$, and the translationally invariant, 
nonlocal one-body density matrix describing the distribution of the struck nucleon in the target, $\rho^{K_s}_\alpha$. 

The $NN$ scattering amplitude $\overline M$ can be parameterized in terms of Wolfenstein
amplitudes~\cite{wolfenstein-ashkin,Fachruddin:2000wv,Golak:2010wz},
\begin{eqnarray}
\label{eq7}
	\overline{M}(\bm{q},\bm{\mathcal{K}}_{NN},\epsilon)&=& A(\bm{q},\bm{\mathcal{K}}_{NN},\epsilon)\textbf{1}\otimes\textbf{1} \nonumber \\
	&+& iC(\bm{q},\bm{\mathcal{K}}_{NN},\epsilon)~\left(\bm{\sigma^{(0)}} \cdot \hat{\bm{n}} \right)\otimes\textbf{1} \nonumber \\
	&+& iC(\bm{q},\bm{\mathcal{K}}_{NN},\epsilon)~\textbf{1}\otimes\left(\bm{\sigma^{(i)}}\cdot \hat{\bm{n}} \right) \nonumber \\
	&+& M(\bm{q},\bm{\mathcal{K}}_{NN},\epsilon)(\bm{\sigma^{(0)}}\cdot\hat{\bm{n}})\otimes(\bm{\sigma^{(i)}}\cdot\hat{\bm{n}}) \nonumber \\
	&+& \left[G(\bm{q},\bm{\mathcal{K}}_{NN},\epsilon)-H(\bm{q},\bm{\mathcal{K}}_{NN},\epsilon)\right](\bm{\sigma^{(0)}}\cdot\hat{\bm{q}})\otimes(\bm{\sigma^{(i)}}\cdot\hat{\bm{q}}) \cr
	&+& \left[G(\bm{q},\bm{\mathcal{K}}_{NN},\epsilon)+H(\bm{q},\bm{\mathcal{K}}_{NN},\epsilon)\right](\bm{\sigma^{(0)}}\cdot\hat{\bm{\mathcal{K}}})\otimes(\bm{\sigma^{(i)}}\cdot\hat{\bm{\mathcal{K}}}) \cr
	&+& D(\bm{q},\bm{\mathcal{K}}_{NN},\epsilon)\left[(\bm{\sigma^{(0)}}\cdot\hat{\bm{q}})\otimes(\bm{\sigma^{(i)}}\cdot\hat{\bm{\mathcal{K}}})+(\bm{\sigma^{(0)}}\cdot\hat{\bm{\mathcal{K}}})\otimes(\bm{\sigma^{(i)}}\cdot\hat{\bm{q}})\right]~,
\end{eqnarray}
where $\bm{\sigma^{(0)}}$ describes the spin of the projectile, and $\bm{\sigma^{(i)}}$ the spin 
of the struck nucleon. The average momentum in the $NN$ frame is defined as $\bm{\mathcal{K}}_{NN} =
\frac{1}{2} \left(\bm{k'}_{NN} + \bm{k}_{NN}\right)$. The scalar functions $A$, $C$, $M$, $G$, $H$, and $D$ are referred to as
Wolfenstein amplitudes and only depend on the scattering momenta and energy. Each term in Eq.~(\ref{eq7}) has two components, namely the scalar function and the coupling between the operators of the projectile and the struck nucleon, in that respective order.
The linear independent unit vectors $\hat{\bm{q}}$, $\hat{\bm{\mathcal{K}}}$, and $\hat{\bm{n}}$ are defined in
terms of the momentum transfer and the average momentum as
\begin{eqnarray}
\hat{\bm{q}}=\frac{\bm{q}}{\left| \bm{q} \right|}~~,~~~ 
\hat{\bm{\mathcal{K}}}=\frac{\bm{\mathcal{K}}}{\left| \bm{\mathcal{K}} \right|} ~~,~~~
 \hat{\bm{n}}=\frac{\bm{\mathcal{K}} \times \bm{q}}{\left| \bm{\mathcal{K}} \times \bm{q}
\right|},
\label{eq8}
\end{eqnarray}
and span the momentum vector space. Again, with the exception of the momentum transfer $\bm q$, which is
invariant under frame transformation, the vectors in Eq.~(\ref{eq8}) need to be considered in their
respective frame. 

All Wolfenstein amplitudes need to be considered when evaluating the effective
interaction in Eq.~(\ref{eq5}). 
For the struck target nucleon the expectation values of the operator ${\bf 1}$ and the scalar products of $\bm{\sigma^{(i)}}$ with the linear independent unit vectors of Eq.~(\ref{eq7}) need to be evaluated with the ground state wave functions. Evaluating the expectation value of the operator ${\bf 1}$ in the ground state of 
the nucleus results in the scalar nonlocal one-body density that has traditionally been used as input to microscopic or {\it ab initio} calculations of leading order effective interactions~\cite{Burrows:2018ggt,Gennari:2017yez,Elster:1996xh,Elster:1989en}. The other operators from Eq.~(\ref{eq7}), namely $(\bm{\sigma^{(i)}} \cdot \hat{\bm n})$, $(\bm{\sigma^{(i)}} \cdot \hat{\bm q})$, and $(\bm{\sigma^{(i)}} \cdot \hat{\bm {\mathcal{K}}})$ have, to our knowledge, not yet been considered in a systematic fashion together with realistic nuclear structure calculations. Only within the framework of a toy model~\cite{Orazbayev:2013dua} such an attempt was made.

To begin, we start with the general expression of nonlocal density as previously described in Ref.~\cite{Burrows:2017wqn} but include the spin operator $\bm{\sigma^{(i)}}$ explicitly,
\begin{eqnarray}
\label{density}
\rho^{K_s}_{q_s}\left(\bm{p}, \bm{p}' \right) = \left\langle \Phi' \left| \sum_{i=1}^{A} \delta^3(
\bm{p_i} - \bm{p}) \delta^3( \bm{p_i}' - \bm{p}') \sigma_{q_s}^{(i) K_s} \right| \Phi \right\rangle~,
\end{eqnarray}
where $\sigma^{(i) K_s}_{q_s}$ is the spherical representation of the spin operator and the wavefunction $\Phi$ $(\bm{p_1}, ..., \bm{p_A}) = \left\langle \bm{p_1}, ..., \bm{p_A} | \Phi \right\rangle$ is defined in momentum space. This nonlocal density, after defining $\sigma^{(i) K_s}_{q_s}$ using Eq.~(\ref{sigma}), can be evaluated using $K_s=0$ to become the nonlocal one-body scalar density that has been used in traditional calculations or using $K_s=1$ to become a nonlocal one-body spin density which up to now has not been evaluated.

Without loss of generality, we choose to present the derivation of the expectation value of the term $(\bm{\sigma^{(i)}} \cdot \hat{\bm{n}})$ explicitly. We define a scalar function $S_n\left(\bm{p},\bm{p}' \right)$ using Eq.~(\ref{density}) as
\begin{eqnarray}
\label{eq10}
	S_n\left(\bm{p},\bm{p}' \right) \equiv \rho^{K_s} \left(\bm{p}, \bm{p}' \right) \cdot \hat{\bm{n}} = \sum_{q_s}{(-1)^{q_s}
	\rho^{K_s=1}_{q_s} \left(\bm{p}, \bm{p}'\right)  \hat{\bm n}^1_{-q_s}},
	\end{eqnarray}
where $K_s=1$ due to the tensor coupling. The scalar product of $(\bm{\sigma_i} \cdot \hat{\bm{n}})$ is, in principle, inside the bra-ket of Eq.~(\ref{density}). When defining $S_n(\bm{p},\bm{p}')$, the vector $\hat{\bm{n}}$ can be moved outside the bra-ket since it only depends on $\bm{p}$ and $\bm{p'}$. The scalar function $S_n(\bm{p},\bm{p}')$ will be from here on referred to as the intrinsic spin-projected momentum distribution.

	Following the same procedure as laid out in Ref.~\cite{Burrows:2017wqn}, we use the Wigner-Eckart theorem, decouple the orbital angular momentum $l$ and the spin $s$ instead of using the total angular momentum $j$. Then the reduced matrix elements are evaluated. This procedure guarantees that in the calculation of the reduced matrix element of the spin operator is included explicitly as
\begin{eqnarray}
\label{eq11}
	\lla s' \left| \left| \sigma^{K_s} \right| \right| s \rra = \hat{s} \hat{K_s} \delta_{s' s},
\end{eqnarray}
with $s'=s=1/2$, $\hat{s}= \sqrt{2s+1}$, and $\hat{K}_s= \sqrt{2K_s+1}$. The translationally invariant one-body density is obtained by using the Talmi-Moshinsky transformation from the ($\bm{p},\bm{p'}$) variables to the ($\bm{q},\bm{\mathcal{K}}$) variables and removing the center-of-mass motion of the nucleus.  The scalar product of the density with $\hat{\bm{n}}$, written in terms of spherical harmonics ($\hat{\bm{n}}_\alpha=\left| \bm{n} \right| \sqrt{4\pi/3} Y^1_\alpha(\hat{\bm n})$) leads to the expression for the expectation value of $(\bm{\sigma^{(i)}} \cdot \hat{\bm{n}})$ in the ground state of the nucleus,
\begin{eqnarray}
\label{eq12}
	S_{n}(\bm{q},\bm{\mathcal{K}}) 
	&=& \sum_{q_s} (-1)^{-q_s} \sqrt{\frac{4\pi}{3}} Y^1_{-q_s}(\hat{\bm n}) \sum_{nljn'l'j'} \sum_{K_l=|l-l'|}^{l+l'} \sum_{k_l=-K_l}^{K_l} \sum_{Kk} \lla K_l k_l 1 q_s | K k \rra \cr
& &(-1)^{J-M} \threej{J}{K}{J}{-M}{k}{M} (-1)^{-l} \hat{j}\hat{j'} \hat{s} \hat{1} \hat{K_l} \ninej{l'}{l}{K_l}{s}{s}{1}{j'}{j}{K} (-i)^{l+l'} \cr
& & \sum_{n_q,n_{\mathcal{K}},l_q,l_{\mathcal{K}}} \lla n_{\mathcal{K}} l_{\mathcal{K}}, n_q l_q : K_l | n' l', n l : K_l \rra_{d=1} R_{n_{\mathcal{K}} l_{\mathcal{K}}}(\mathcal{K}) R_{n_q l_q}(q) \mathcal{Y}_{K_l k_l}^{*l_q l_{\mathcal{K}}}(\widehat{\bm q},\hat{\bm{\mathcal{K}}}) \cr
& &\lla A \lambda J \left|\left| (a^{\dagger}_{n'l'j'} \tilde{a}_{nlj})^{(K)} \right|\right| A \lambda J \rra e^{\frac{1}{4A} b^2 q^2},
\end{eqnarray}
where the term $e^{\frac{1}{4A} b^2 q^2}$ stems from the removal of the center of mass motion of the nucleus. The notation used in Eq.~(\ref{eq12}) and the removal of the center-of-mass motion is explicitly derived in Ref.~\cite{Burrows:2017wqn}. The use of the radial harmonic oscillator wave-functions $R_{n_q l_q}$ and harmonic oscillator length parameter $b=\sqrt{\frac{\hbar^2c^2}{mc^2\hbar\omega}}$ indicates a use of the harmonic oscillator basis in the NCSM calculation. By choosing the vector $\hat {\bm q}$ in the $z$-direction and $\hat{\bm{\mathcal{K}}}$ in the $x$-$z$ plane,
the direction of $\hat{\bm{n}}$ is in the negative $y$-direction. Since the ground state of a $0^+$ nucleus is in a
state of angular momentum $J = 0$, the expression can be further simplified to
\begin{eqnarray}
\label{eq13}
	S_{n}(\bm{q},\bm{\mathcal{K}}) &=& (-i) \sqrt{3} \sum_{nljn'l'j'} 
 (-1)^{-l} \hat{j}\hat{j} \ninej{l'}{l}{1}{\frac{1}{2}}{\frac{1}{2}}{1}{j}{j}{0} (-i)^{l+l'} \cr
& & \sum_{n_q,n_{\mathcal{K}},l_q,l_{\mathcal{K}}} \lla n_{\mathcal{K}} l_{\mathcal{K}}, n_q l_q : 1 | n' l', n l : 1 \rra_{d=1} R_{n_{\mathcal{K}} l_{\mathcal{K}}}(\mathcal{K}) R_{n_q l_q}(q) \sum_{q_s=-1,1} \mathcal{Y}_{1 -q_s}^{*l_q l_{\mathcal{K}}}(\widehat{\bm q},\hat{\bm{\mathcal{K}}}) \cr
& &\lla A \lambda 0 \left|\left| (a^{\dagger}_{n'l'j'} \tilde{a}_{nlj})^{(0)} \right|\right| A \lambda 0 \rra e^{\frac{1}{4A} b^2 q^2}~.
\end{eqnarray}
The scalar function $S_n(\bm{q},\bm{\mathcal{K}})$ represents the expectation value of the
spin operator projected along $\hat{\bm{n}}$ in the ground state of the nucleus.  
More details of its derivation are given in
Appendix~\ref{appendixA} and Ref.~\cite{Popa:2019iwx}.

The expectation values of the remaining scalar products, $(\bm{\sigma^{(i)}} \cdot
\hat{\bm q})$ and $(\bm{\sigma^{(i)}} \cdot \hat{\bm {\mathcal{K}}})$, can be derived in a
similar fashion, leading to functions $S_q(\bm{q},\bm{\mathcal{K}})$ and
$S_\mathcal{K}(\bm{q},\bm{\mathcal{K}})$. However, considering the scalar products more
closely, $(\bm{\sigma^{(i)}} \cdot \hat{\bm q})$ represents a scalar product of a pseudo-vector
with a vector, a construct that is not invariant under parity transformations, and thus should
not contribute to the effective interaction. We verified that this is indeed the case by
explicitly calculating that the expectation value $S_q(\bm{q},\bm{\mathcal{K}})$ is zero. The same
is true for the expectation value $(\bm{\sigma^{(i)}} \cdot \hat{\bm {\mathcal{K}}})$, which
also gives a zero contribution in the ground state. Therefore, none of the Wolfenstein amplitudes $G$, $H$, and $D$ contribute to the $NA$ elastic scattering amplitude.

After evaluating the expectation values of the scalar products of the spin of the
struck target nucleon with all three momentum vectors, and realizing that only the expectation
value of $\left(\bm{\sigma^{(i)}}\cdot\hat{\bm{n}} \right)$ leads to a non-vanishing contribution, we know
that only
the first four terms of the $NN$ scattering amplitude as written in Eq.~(\ref{eq7}) 
contribute to the effective interaction $ \widehat{U}(\bm{q},\bm{\mathcal{K}_{NA}},\epsilon)$
from Eq.~(\ref{eq5}). The first two of them, Wolfenstein amplitudes $A$ and $C$ traditionally
correspond to the central and spin-orbit parts of the effective interaction. Taking into
account the spin of the struck nucleon leads to additional contributions. The spin-orbit term
$iC(\bm{q},\bm{\mathcal{K}}_{NN},\epsilon)~\textbf{1}\otimes\left(\bm{\sigma^{(i)}}\cdot
\hat{\bm{n}} \right)$ of Eq.~(\ref{eq7}) leads to a modification of the central part of the $NA$ effective potential,
whereas the term
$M(\bm{q},\bm{\mathcal{K}}_{NN},\epsilon)(\bm{\sigma^{(0)}}\cdot\hat{\bm{n}})\otimes(\bm{\sigma^{(i)}}\cdot\hat{\bm{n}})$
contributes to the spin-orbit part  of the $NA$ effective potential.
In order to calculate the quantity in Eq.~(\ref{eq5}), we need to combine the Wolfenstein amplitudes in Eq.~(\ref{eq7}) with the density defined in Eq.~(\ref{density}) projected along the relevant vectors. 
Thus, the effective interaction of Eq.~(\ref{eq5}) between e.g. a proton and a nucleus
 is explicitly written as
\begin{eqnarray}
\label{SymbolicU}
\lefteqn{\widehat{U}(\bm{q},\bm{\mathcal{K}}_{NA},\epsilon) =} & &  \cr
& & \sum_{\alpha=n,p} \int d^3{\mathcal{K}} \eta\left( \bm{q}, \bm{\mathcal{K}}, \bm{\mathcal{K}}_{NA} \right) 
A_{p\alpha}\left( \bm{q}, \frac{1}{2}\left( \frac{A+1}{A}\bm{\mathcal{K}}_{NA} - \bm{\mathcal{K}} \right); \epsilon
\right) \rho_\alpha^{K_s=0} \left(\bm{\mathcal{P}'}, \bm{\mathcal{P}}  \right) \cr  
&+&i (\bm{\sigma^{(0)}}\cdot\hat{\bm{n}}) \sum_{\alpha=n,p} \int d^3{\mathcal{K}} \eta\left( \bm{q},
\bm{\mathcal{K}}, \bm{\mathcal{K}}_{NA} \right) 
C_{p\alpha}\left( \bm{q}, \frac{1}{2}\left( \frac{A+1}{A}\bm{\mathcal{K}}_{NA} - \bm{\mathcal{K}} \right);
\epsilon
\right) \rho_\alpha^{K_s=0} \left(\bm{\mathcal{P}'}, \bm{\mathcal{P}}  \right) \nonumber \cr 
&+&i \sum_{\alpha=n,p} \int d^3{\mathcal{K}} \eta\left( \bm{q}, \bm{\mathcal{K}}, \bm{\mathcal{K}}_{NA}
\right) C_{p\alpha} \left( \bm{q}, \frac{1}{2}\left( \frac{A+1}{A}\bm{\mathcal{K}}_{NA} - \bm{\mathcal{K}}
\right); \epsilon \right) S_{n,\alpha} \left(\bm{\mathcal{P}'}, \bm{\mathcal{P}} \right) \cos \beta\cr
&+&i (\bm{\sigma^{(0)}}\cdot\hat{\bm{n}}) \sum_{\alpha=n,p} \int d^3{\mathcal{K}} \eta\left( \bm{q},
\bm{\mathcal{K}}, \bm{\mathcal{K}}_{NA} \right)  (-i) 
M_{p\alpha} \left( \bm{q}, \frac{1}{2}\left( \frac{A+1}{A}\bm{\mathcal{K}}_{NA} - \bm{\mathcal{K}}
\right); \epsilon \right) S_{n,\alpha} \left(\bm{\mathcal{P}'}, \bm{\mathcal{P}}  \right) \cos \beta  , 
\end{eqnarray}
\noindent
where $\bm{\mathcal{P}'}=\left(\bm{\mathcal{K}} - \frac{A-1}{A} \frac{\bm{q}}{2}\right)$ and
$\bm{\mathcal{P}}=\left(\bm{\mathcal{K}} + \frac{A-1}{A} \frac{\bm{q}}{2}\right)$.
The quantity $\rho_{\alpha}^{K_s=0}$, is the scalar density  derived in Ref.~\cite{Burrows:2018ggt} and 
 $S_{n,\alpha}$ is given in Eq.~(\ref{eq13}). The term $i (\bm{\sigma^{(0)}}\cdot\hat{\bm{n}})$ represents
the `usual' spin-orbit operator in momentum space.  The above expression clearly shows how taking into account the
spin of the struck nucleon adds a term to the central as well as the spin-orbit part of the effective
interaction. 

The last two terms in Eq.~(\ref{SymbolicU}) show a factor
$\cos \beta $, which represents the frame transformation $\textbf{1}\otimes\left(\bm{\sigma^{(i)}}\cdot
\hat{\bm{n}} \right)$ between the frame of the target nucleus and   
the $NN$ frame. It is
necessary to take this transformation into account, since the Wolfenstein amplitudes are calculated via
the solution of a LS equation in the $NN$ center-of-mass (c.m.) frame  with a given $NN$ potential.
In that calculation the unit vector ${\hat n}$ is defined in the $NN$ frame. The function
$S_n\left(\bm{p},\bm{p}' \right)$ from Eq.~(\ref{eq10}) is calculated in the
frame of the target nucleus. Thus only the component of $\hat{\bm{n}}$ projected on the normal
of the $NN$ scattering frame, ${\hat{\bm{n}}}_{NN}$ will contribute to the effective interaction.
We therefore define the scalar product 
$\hat{\bm{n}} \cdot {\hat{\bm{n}}}_{NN} \equiv \cos \beta$, which will project the normal vector
of the nucleus frame to the $NN$ frame, with $\beta$ given as
\begin{eqnarray}
\label{eq14}
	\cos \beta = \cos(\phi - \phi_{NN}) = \cos\left(\phi - \tan^{-1}\left( \frac{-\mathcal{K}\sin(\theta)\sin(\phi)}{\frac{A+1}{A} \mathcal{K}_{NA}\sin(\theta_{NA}) - \mathcal{K}\sin(\theta)\cos(\phi)} \right) \right).
\end{eqnarray}
The explicit derivation of $\cos \beta$ is described in Appendix~\ref{appendixB}.


\section{Results and Discussion}
\label{results}

In this Section we present calculations of observables for elastic scattering from closed as well as
open-shell nuclei in which the leading order effective interaction is calculated {\it ab initio}, i.e. the
$NN$ interaction is taken into account consistently in the structure as well as reaction calculation. For the reaction calculation the $NN$ amplitudes are represented in form of Wolfenstein amplitudes, Eq.~(\ref{eq7}). We are considering elastic scattering off $0^+$ nuclei. As discussed in Sec.~\ref{formal}, the spin-projections
of the struck target nucleon with the vectors $\bm{q}$ and $\bm{\mathcal{K}}$ are zero, leaving only the Wolfenstein amplitudes $A$, $C$, and $M$ contributing to the effective $NA$ interaction, representing scalar, vector, and tensor components of the $NN$ interaction. 

Traditional calculations of the leading order
term~\cite{Elster:1996xh,Chinn:1994xz,Burrows:2018ggt,Gennari:2017yez}, despite using realistic one-body
densities, neglected the spin of the struck target nucleon, and concentrated  on closed-shell
nuclei, arguing that for closed-shell nuclei those spin contributions most likely average out.
If the spin of the struck target nucleon is ignored, one can immediately see from Eq.~(\ref{eq7}) that
only the Wolfenstein amplitudes $A$ and $C$ contribute, leading to the traditional central and spin-orbit 
parts of the effective interaction. 

In the following sections we will inspect the effect of ignoring the spin of the struck target nucleon in the
effective interaction for closed-shell and open-shell nuclei on elastic scattering observables at
projectile kinetic energies where the leading order term in the multiple scattering expansion should dominate. 
We will also examine scattering observables at lower energies, which are somewhat outside the validity realm of the leading order, to study the energy dependence of the  effective interaction
compared to the approximation in which the spin of the struck target nucleon is ignored. 

\begin{figure}
\centering
\includegraphics[width=12cm]{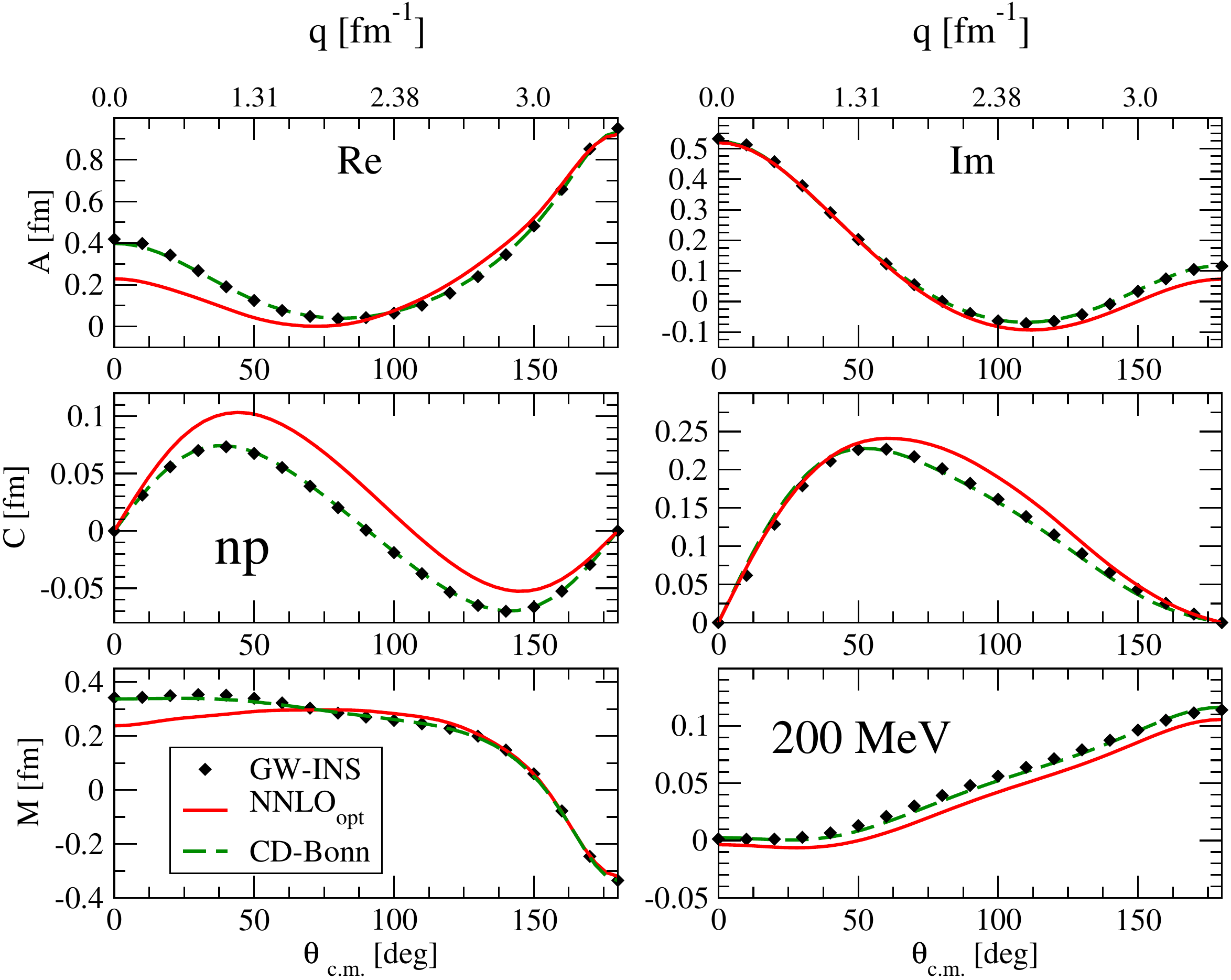}
\caption{Wolfenstein amplitudes A and C as function of the scatting angle and
momentum transfer for $np$ scattering at 200~MeV laboratory kinetic energy. The
solid (red) line represents  the NNLO${_{\rm{opt}}}$ chiral
interaction~\protect\cite{Ekstrom13}, and the dashed (green) line  the CD-Bonn
potential~\protect\cite{Machleidt:2000ge}. 
The solid diamonds stand for the extraction from the
GW-INS analysis~\protect\cite{Workman:2016ysf}.
}
\label{fig1}
\end{figure}

\begin{figure}
\centering
\includegraphics[width=12cm]{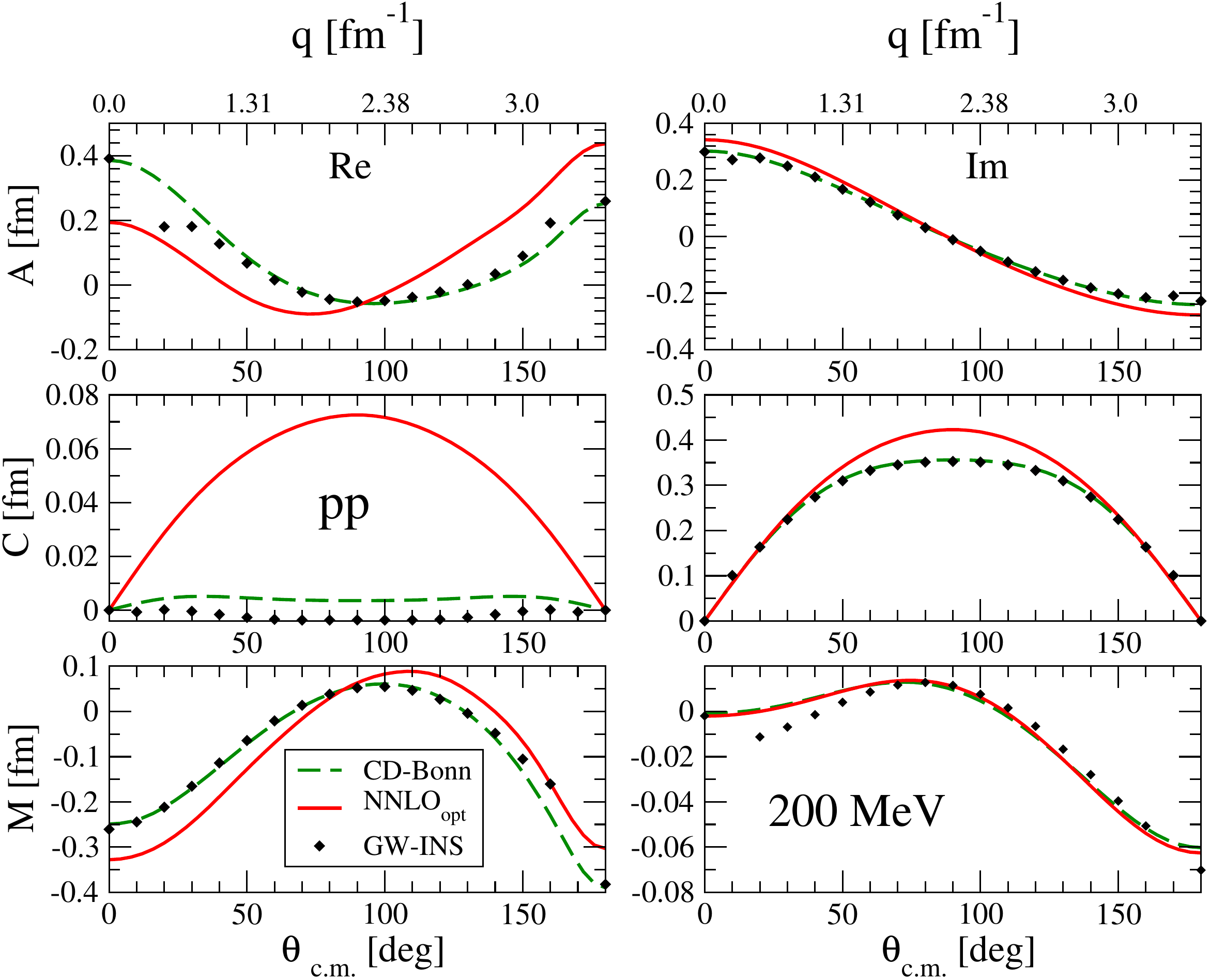}
\caption{Same as Fig.~\ref{fig1} for $pp$ scattering at 200~MeV laboratory kinetic energy.
}
\label{fig2}
\end{figure}

For the calculations of the scalar and spin projected one-body densities as well as the $NN$
scattering amplitudes we choose the optimized chiral $NN$ interaction at the
next-to-next-to-leading order NNLO$_{\rm{opt}}$ from Ref.~\cite{Ekstrom13}. This interaction
is fitted  for $NN$ laboratory energies up to 125 MeV.
In the A$~=3$ and A$~=4$ systems the contributions of the $3NF$s are smaller than in most other
parameterizations of chiral interactions. 
As a consequence, nuclear quantities like root-mean-square radii
and electromagnetic transitions in light and intermediate-mass nuclei can be calculated
reasonably well without invoking $3NF$s \cite{Henderson:2017dqc,Dytrych:2020vkl,Baker:2020rbq}. Since $NA$
scattering calculations discussed here concentrate on the energy regime between about 100 and 200~MeV, we
will have to employ this interaction beyond its fitted energy range. The authors of
Ref.~\cite{Ekstrom13} give a $\chi^2/{\rm datum} \approx 2$ for $np$ scattering between 125 and 183~MeV and
$\approx24$ for $pp$ scattering.  In
Figs.~\ref{fig1} and~\ref{fig2} we show the Wolfenstein amplitudes $A$, $C$, and $M$ for
$np$ and $pp$ scattering at 200~MeV laboratory kinetic energy together with the experimental
extraction from the GW-INS analysis~\cite{Workman:2016ysf}. To compare, we also show those
Wolfenstein amplitudes obtained from the Charge-Dependent Bonn potential (CD-Bonn)~\cite{Machleidt:2000ge},
which is fitted to the $NN$ data up to 300~MeV with $\chi^2/{\rm datum}\approx 1$. 
The largest deviations from the experimental extraction occurs for the amplitude $C$, which
for $np$ scattering is moderately over-predicted, while for $pp$ scattering the real part of
the amplitude is severely over-predicted. This is consistent with remarks in Ref.~\cite{Ekstrom13} that
the $NN$ p-waves are less well represented. The $NN$ spin-orbit force is dominated by p-waves and manifests
itself in the Wolfenstein amplitude $C$. It is interesting to note that while relaxing the fit to $NN$
data, the interaction appears to effectively include spin effects that otherwise enter through the $3NF$s, which may be the reason for the reasonably good descriptions of spin observables in light nuclei as shown in Ref.~\cite{Burrows:2018ggt} and in the present outcomes. 
The Wolfenstein amplitude $M$ captures contributions of the tensor part of the $NN$ interaction, which is
quite well represented by the NNLO$_{\rm opt}$ chiral interaction.

At 100~MeV the description of the same Wolfenstein amplitudes by the NNLO$_{\rm opt}$ chiral interaction
is much better, since this energy is still within the regime where the interaction is fitted with a much
smaller $\chi^2/{\rm datum}$. However, even at 100~MeV, the amplitude $C$ is still slightly over-predicted.
 Corresponding figures can be found in the supplemental material.

\subsection{Closed-shell nuclei $^4$He and $^{16}$O}

The most natural question is how the scattering observables for closed-shell nuclei are affected by
neglecting the spin of the struck target nucleon, having in mind that this approximation has always been
employed. 
Thus, comparing the {\it ab initio} leading order calculation with the traditionally employed
approximation should answer the question whether  ignoring the spin of the struck nucleon in this case
was reasonable.

\begin{figure}
\centering
\includegraphics[width=12cm]{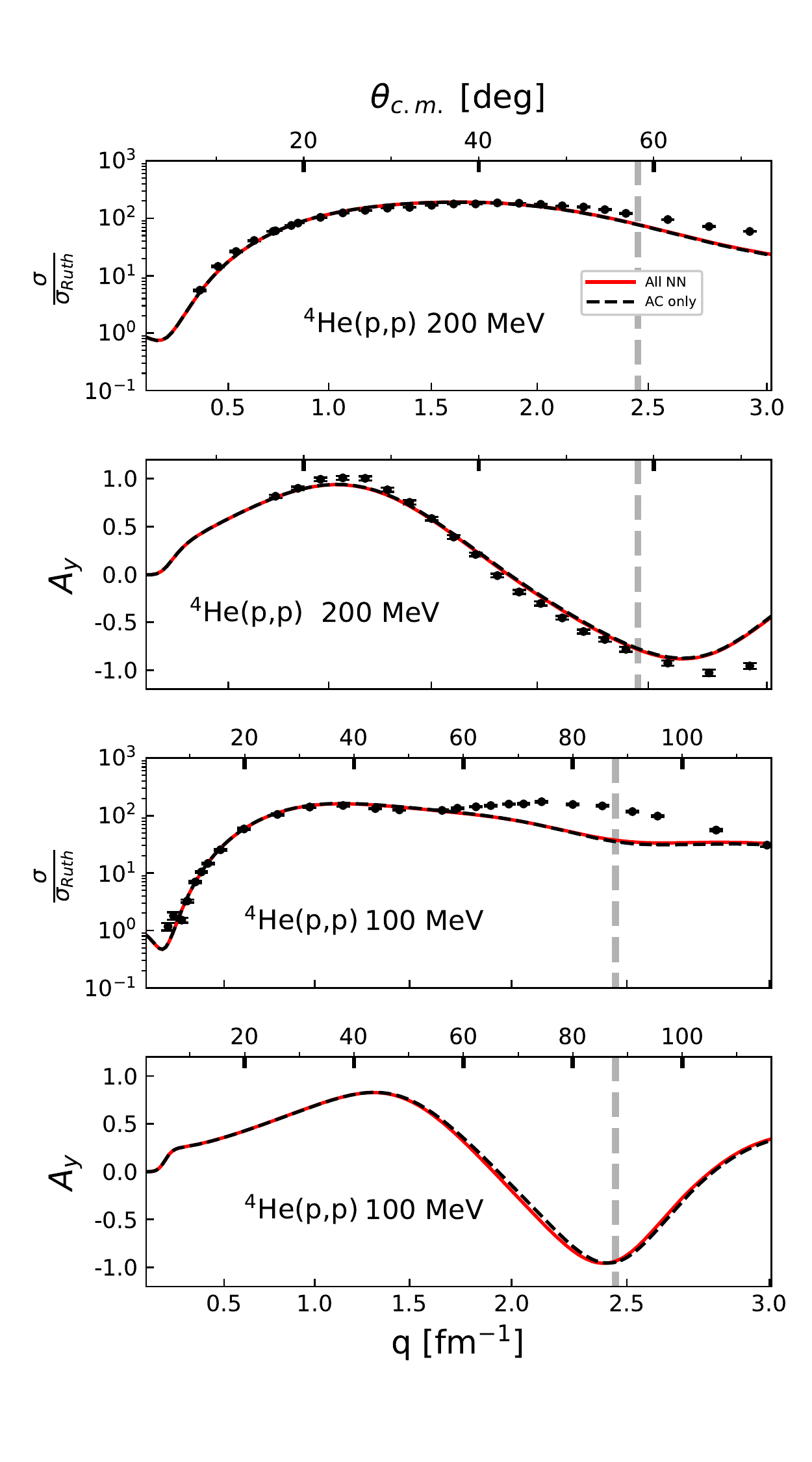}
\caption{The angular distribution of the differential cross section divided by the Rutherford cross 
section and the analyzing power for elastic proton scattering from $^4$He at 200 and 100~MeV
laboratory kinetic energy as a function of the momentum transfer and the c.m. angle calculated with the
NNLO$_{\rm opt}$ chiral interaction~\protect\cite{Ekstrom13}. The solid (red) line represents the 
calculations with the full \textit{NN} interaction, while for the calculations represented by the 
dashed (black) line the 
 the spin of the struck nucleon in the target is neglected. Both calculations employ
$N_{\rm max}$=18 and $\hbar \omega$=20. The data for 200~MeV are taken
from Ref.~\cite{Moss:1979aw}, and for 100~MeV from Ref.~\cite{Goldstein:1970dg}. The dashed vertical
line in each figure indicates the momentum transfer $q=2.45$~fm$^{-1}$ corresponding to the
laboratory kinetic energy of 125~MeV of the \textit{NN} system.
}
\label{fig3}
\end{figure}

In Fig.~\ref{fig3} both,  the angular distribution of the differential cross section divided by the
Rutherford cross section as well as the analyzing power for elastic scattering of protons off $^4$He is
shown at 200  as well as 100~MeV laboratory projectile kinetic energy. 
The figure  compares the {\it ab initio} calculation, labeled ``All NN", with the traditional approximation
ignoring the spin of the struck target nucleon, labeled ``AC only".
For both calculations we used $N_{\rm max}$=18 and $\hbar\omega$=20~MeV, which is sufficient to obtain converged results to within the plotted line thickness. The grey bar seen in all four figures 
represent the momentum transfer corresponding to the energy range  
of 125 MeV in the $NN$ system, for which  NNLO$_{\rm opt}$ was fitted. 
The figure clearly shows that the spin of the struck nucleon plays an almost imperceptible role in the final result at both projectile energies and in both observables.

\begin{figure}
\centering
\includegraphics[width=12cm]{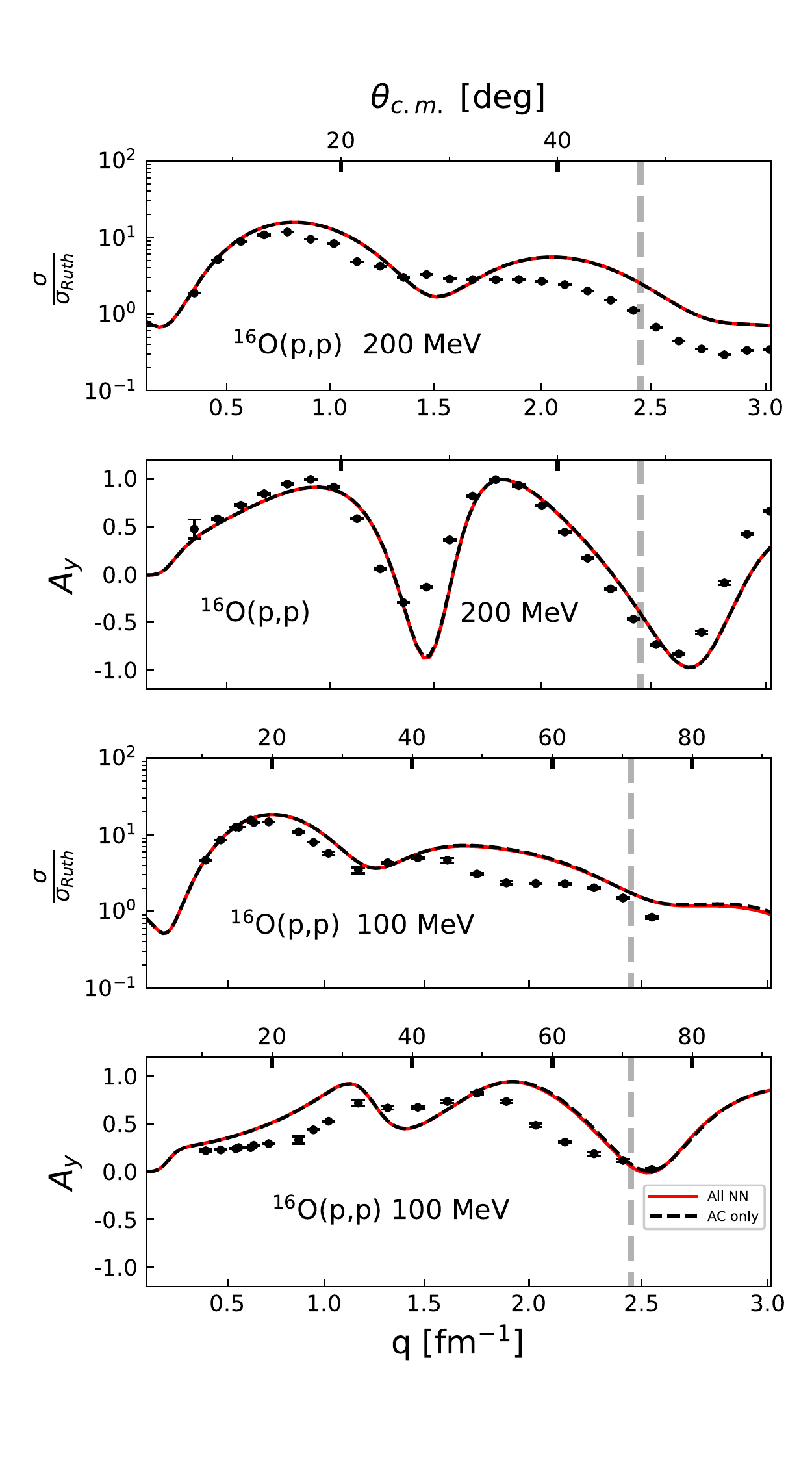}
\caption{The angular distribution of the differential cross section divided by the Rutherford cross
  section and the analyzing power for elastic proton scattering from $^{16}$O at 200 and 100~MeV
  laboratory kinetic energy as a function of the momentum transfer and the c.m. angle calculated with the
  NNLO$_{\rm opt}$ chiral interaction~\protect\cite{Ekstrom13}. The lines follow the same notation as
Fig.~\ref{fig3}. Both calculations employ $N_{\rm max}$=10 and $\hbar \omega$=20. 
The data for 200~MeV are taken from Ref.~\cite{Glover:1985xd}, and for 100~MeV from
Ref.~\cite{Seifert:1990um}.
} 
\label{fig4}
\end{figure}

In Fig.~\ref{fig4}, the same type of comparison is shown for $^{16}$O at 200 and 100~MeV. 
 In the case of $^{16}$O,  $N_{\rm max}$=10 is used, which is not high enough to arrive at a converged result independent of $\hbar\omega$. The spread of the results as they relate to $\hbar\omega$ is given in Ref.~\cite{Burrows:2018ggt}.  In Fig.~\ref{fig4} only the calculations using $\hbar\omega$=20 MeV are shown. The spread due to different values of $\hbar\omega$ is not affected by the inclusion of the spin of the struck nucleon.

For $^{16}$O, NNLO$_{\rm opt}$ gives a significantly smaller charge radius compared to the experimental value:
about 2.39 fm versus  2.70 fm \cite{Angeli:2013epw}. This can be seen in the location of the first minimum 
of the differential cross section in Fig.~\ref{fig4} at both energies. For $^4$He the prediction of the charge
radius  matches more closely, about 1.66~fm compared to the experimental value of 1.68~fm~\cite{Angeli:2013epw}. The same $N_{\rm max}$ and $\hbar\omega$ is used in the charge radii calculations as those in the scattering calculations. This result can partially explain the particularly good description of the calculation as compared to the experimental data.

Both closed shell nuclei, $^4$He and $^{16}$O, lead
to the conclusion that here the spin of the struck nucleon can be neglected in calculating observables for
elastic scattering. 
They confirm that the traditional approximation of ignoring the spin of the struck target
nucleon when considering closed-shell nuclei was justified.

\subsection{Open-shell nuclei $^6$He, $^8$He, and $^{12}$C}

In open-shell nuclei an assumption that spin contributions of the struck target nucleon average out when
summing over all nucleons, is less justified. Thus we examine elastic scattering observables of 
the Helium isotopes, $^6$He and $^8$He, as well as $^{12}$C at 200 and 100~MeV projectile kinetic energy.

\begin{figure}
\centering
\includegraphics[width=12cm]{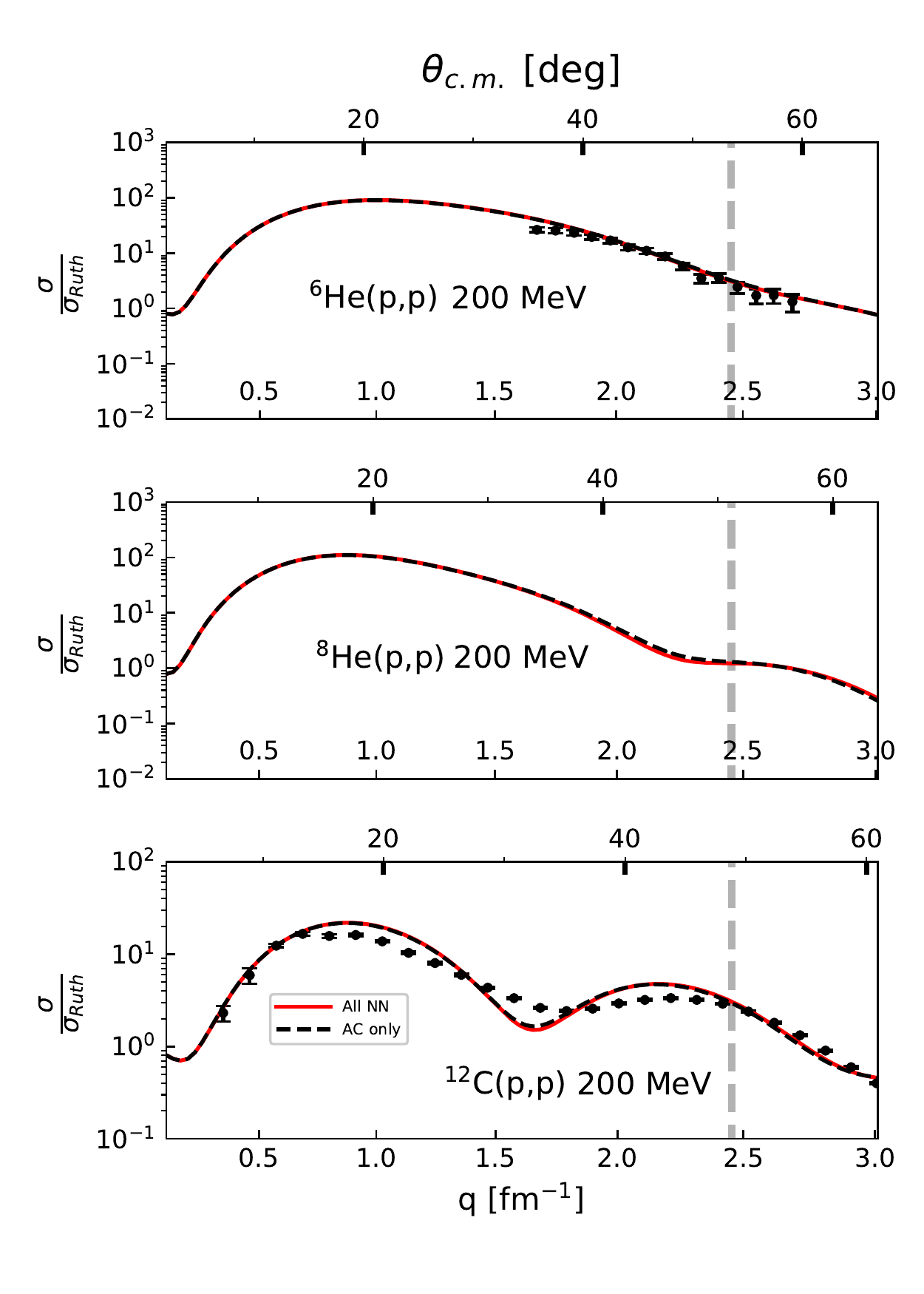}
\caption{The angular distribution of the differential cross section divided by the Rutherford cross
section for elastic proton scattering from $^6$He, $^8$He, and $^{12}$C at 200~MeV laboratory kinetic energy as a function of the momentum transfer and the c.m. angle calculated with the NNLO$_{\rm opt}$ 
chiral interaction~\protect\cite{Ekstrom13}. The lines follow the same notation as
  Fig.~\ref{fig3}. All calculations employ $\hbar \omega$=20 with  $N_{\rm max}$=18 for 
$^6$He, $N_{\rm max}$=14 for $^8$He and $N_{\rm max}$=10 for $^{12}$C.
The data for  $^6$He are taken from Ref.~\cite{Chebotaryov:2018ilv}, 
and for $^{12}$C from Ref.~\cite{Meyer:1981na}.
}
\label{fig5}
\end{figure}

In Fig.~\ref{fig5}, the differential cross sections divided by the Rutherford cross section for elastic scattering of protons off $^6$He, $^8$He, and $^{12}$C are shown for 200 MeV laboratory projectile
energy. The $^6$He
calculations employ  $N_{\rm max}$=18,  while for the $^8$He calculations $N_{\rm max}$=14
 and for the $^{12}$C calculations  $N_{\rm max}$=10 is used. In all cases we use $\hbar\omega$=20 MeV. The dependence of the $^{12}$C calculation on $\hbar\omega$ is shown in detail in Ref.~\cite{Burrows:2018ggt} while the convergence of $^6$He and $^8$He with respect to $\hbar\omega$ is within the line thickness.
The line styles follow that of Fig.~\ref{fig3}.
All three nuclei show almost no difference between the {\it ab initio} calculation and 
the traditional approximation of ignoring the spin of the struck nucleon in the nucleus.

\begin{figure}
\centering
\includegraphics[width=12cm]{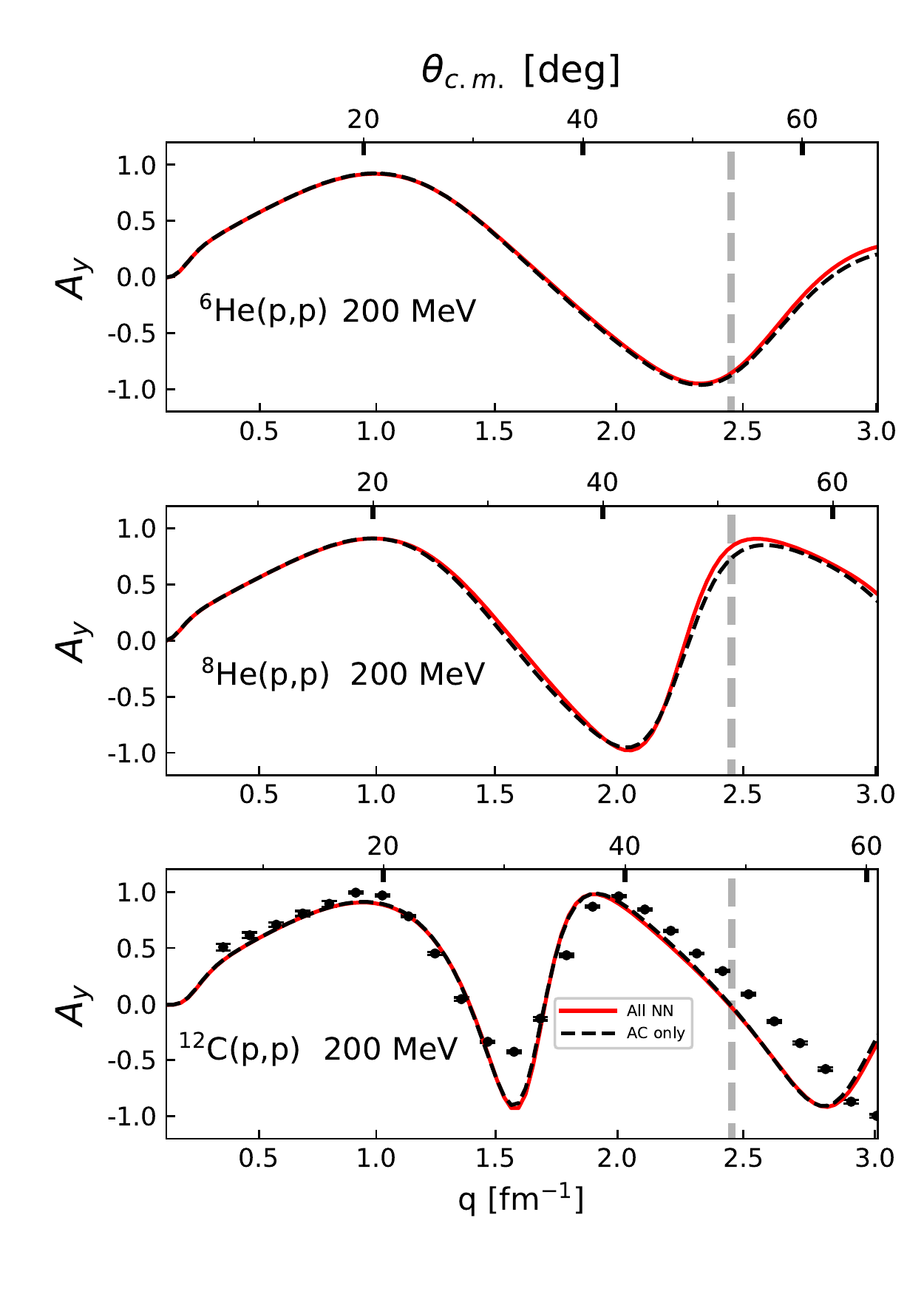}
\caption{The angular distribution of the analyzing powers for elastic proton scattering from $^6$He,
$^8$He, and $^{12}$C at 200~MeV laboratory kinetic energy as a function of the momentum transfer and
the c.m. angle calculated with the NNLO$_{\rm opt}$ chiral interaction~\protect\cite{Ekstrom13}. The lines follow the same notation as Fig.~\ref{fig3}, using the parameters for the structure calculation given in Fig.~\ref{fig5}.  The data for $^{12}$C are taken from Ref.~\cite{Meyer:1981na}.
}
\label{fig6}
\end{figure}

The corresponding analyzing powers for scattering  off $^6$He, $^8$He, and $^{12}$C
are shown in Fig.~\ref{fig6}.
For $^{12}$C, the effect is again negligible. However, for the Helium isotopes, $^6$He and $^8$He, there
is a small but noticeable effect from the spin of the target nucleon at higher momentum transfers. 
The change in the $\hbar\omega$ dependence due to the addition of the spin of the struck nucleon is negligible.

For the predictions of the charge radii of $^6$He, $^8$He, and $^{12}$C, the NNLO$_{\rm opt}$ interaction performs reasonably well. The charge radii predicted for the Helium isotopes $^6$He and $^8$He fall within 6\% for both, 1.95~fm compared to the experimental value of 2.07~fm for $^6$He and 1.90~fm compared to the experimental value of 1.92~fm for $^8$He \cite{Angeli:2013epw}. For $^{12}$C, the predicted charge radius  lies within 5\% using the $\hbar\omega$ value of 20 MeV, namely 2.35~fm compared to the experimental value of 2.47~fm \cite{Angeli:2013epw}.  The same $N_{\rm max}$ and $\hbar\omega$ is used in the calculations of the charge radii and in the scattering calculations. The spread of the calculated values of the charge radius due to the choice of $\hbar\omega$ contains the experimental value. This accuracy can be seen in the analyzing power results at 200 MeV and the very close replication of the dip location around $q=1.5$
fm$^{-1}$. However, the cross section for $^{12}$C is less well described.

\begin{figure}
\centering
\includegraphics[width=12cm]{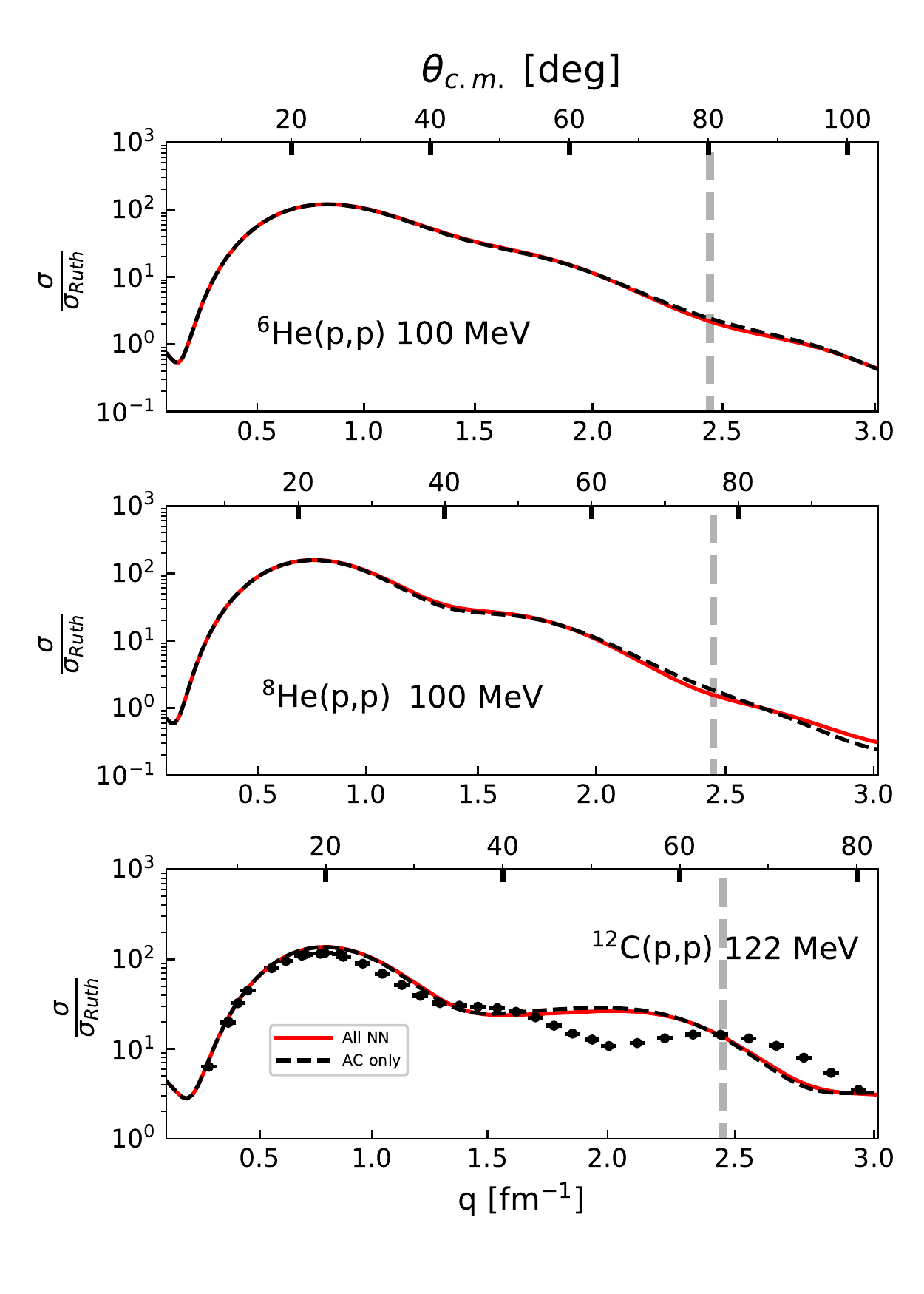}
\caption{Same as Fig.~\ref{fig5} but for 100~MeV projectile kinetic energy in the case of $^6$He and $^8$He,
and 122~MeV projectile kinetic energy for $^{12}$C. The data for $^{12}$C at 122~MeV are taken from
Ref.~\cite{Meyer:1983kd}.
}
\label{fig7}
\end{figure}

\begin{figure}
\centering
\includegraphics[width=12cm]{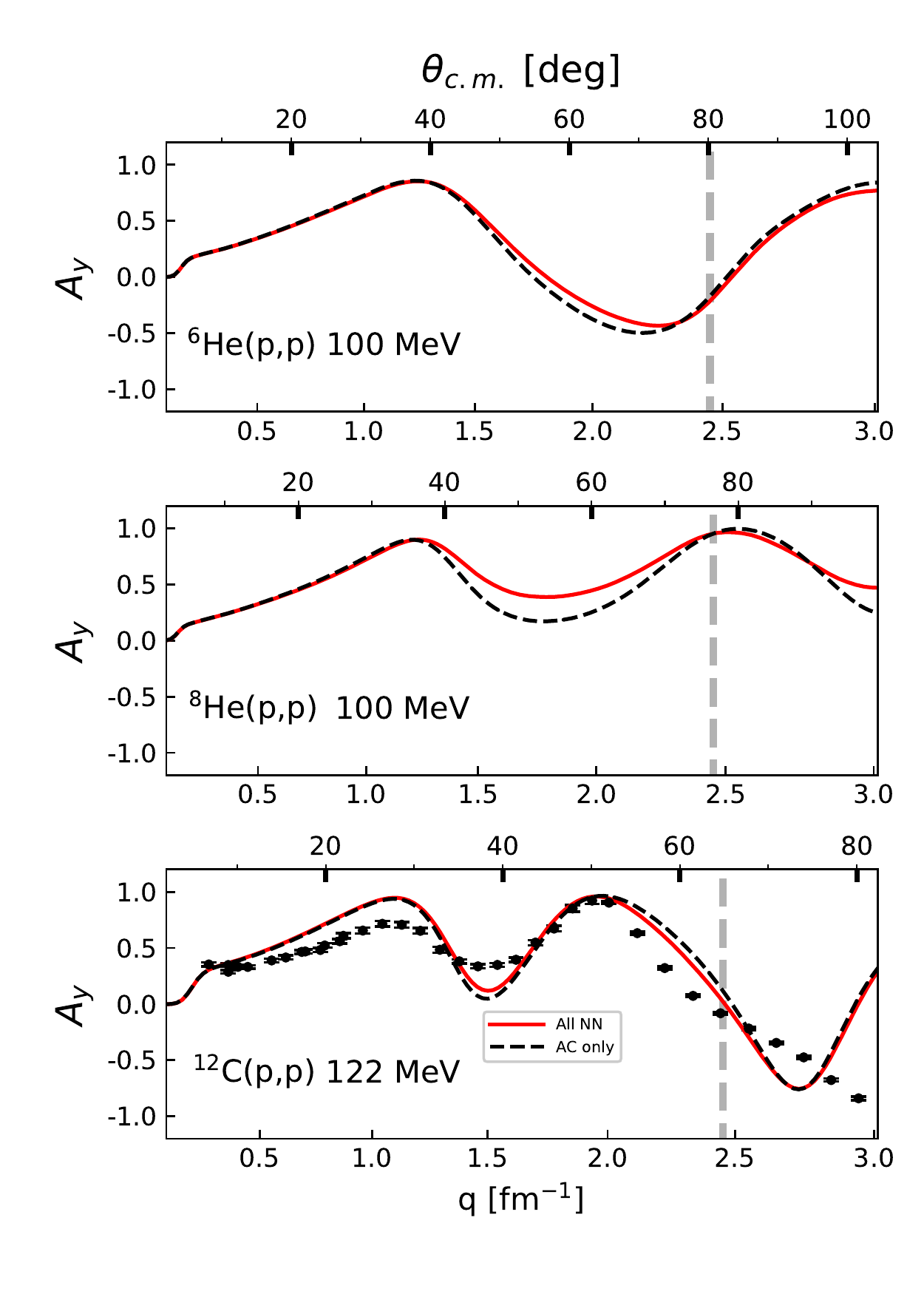}
\caption{Same as Fig.~\ref{fig6}, but for 100~MeV projectile kinetic energy in the case of 
$^6$He and $^8$He, and 122~MeV projectile kinetic energy for $^{12}$C. The data for $^{12}$C at 122~MeV are taken from Ref.~\cite{Meyer:1983kd}.
}
\label{fig8}
\end{figure}

In Fig.~\ref{fig7}, the differential cross section divided by the Rutherford cross section is
shown at 100 MeV projectile kinetic energy for the Helium isotopes and 122~MeV for
$^{12}$C. Again, the cross section is almost unaffected whether the spin of the struck nucleon in the
nucleus is taken into consideration or not.
However, in the analyzing powers calculated at the same energies, Fig.~\ref{fig8}, a difference between the {\it ab initio} calculation and the approximation neglecting the spin of the struck nucleon can be seen.
Both  Helium isotopes show an effect that is larger for $^8$He than $^6$He at this energy. 
For $^{12}$C the difference between the calculations is much smaller, indicating that ignoring the spin of the struck nucleon is also a reasonable approximation in the case of $^{12}$C. This could lead to a  speculation that when considering effective interactions involving nuclei with higher $N/Z$ ratio it becomes more important to take the spin of the struck nucleon in the nucleus into account. However, this will have to be explored with other isotope chains.

\begin{figure}
\centering
\includegraphics[width=12cm]{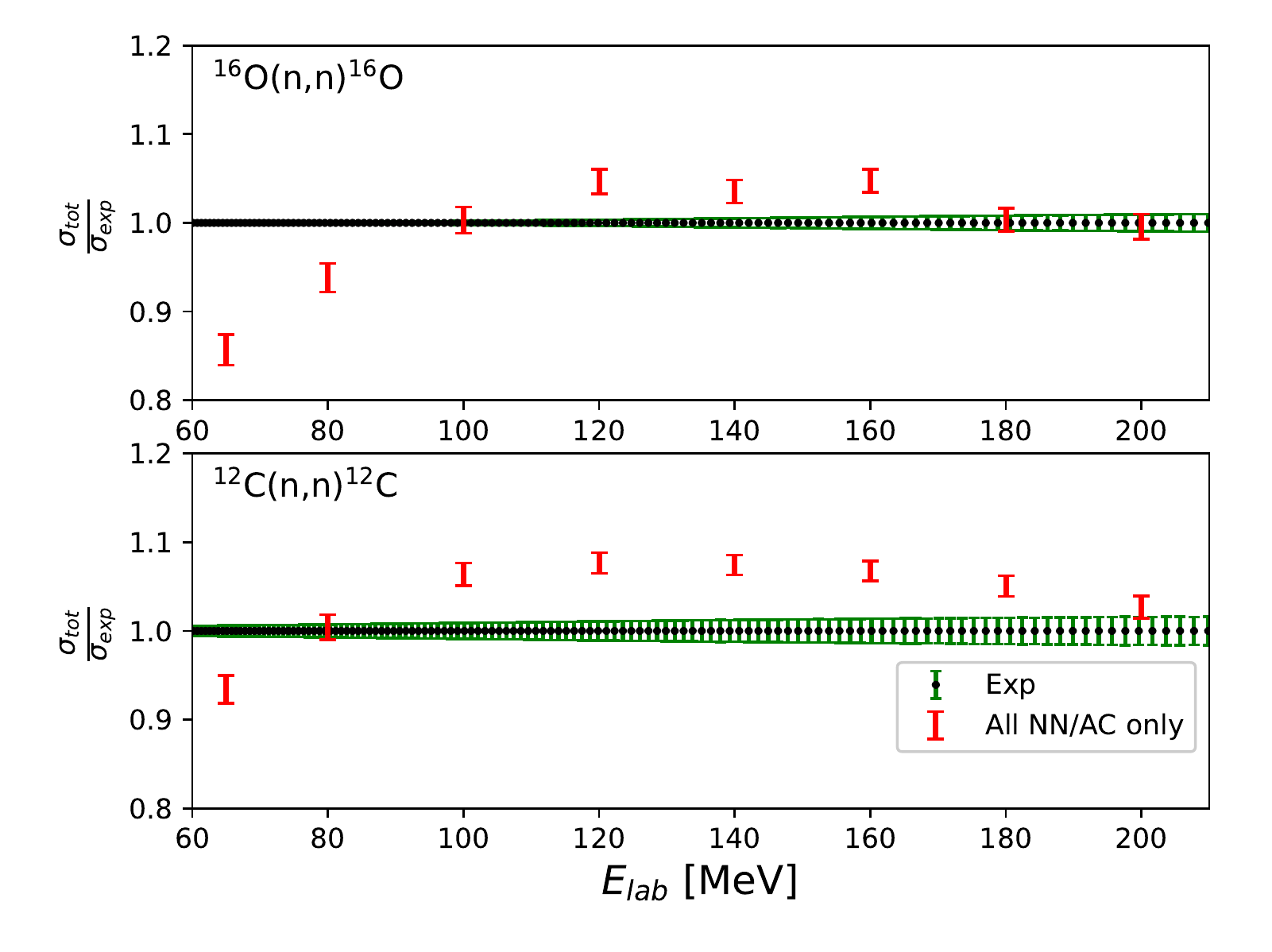}
\caption{The total cross section for neutron scattering from $^{16}$O and $^{12}$C as a function of the
neutron incident energy normalized to the experimental cross section. The solid (red) error bars indicate
the  calculations with the full NN interaction, and coincide with calculations in which the spin of
the struck target nucleon is neglected.
The calculations
use $\hbar \omega$=20 in both cases and go to $N_{\rm max}$=10 for both $^{16}$O and $^{12}$C. The data are taken from Ref.~\cite{Finlay:1993hk}.  
}
\label{fig9}
\end{figure}

Last, we examine 
the total cross sections for neutron scattering off $^{16}$O and $^{12}$C,  shown in Fig.~\ref{fig9}. 
Since the differential cross sections for proton scattering off those nuclei did not show any sensitivity
to the spin of the struck target nucleon, we expect that the total neutron cross section behaves
accordingly. This is indeed the case, as is illustrated in Fig.~\ref{fig7}, where only the {\it ab initio}
calculation is shown, since neglecting the spin of the struck nucleon leads to the almost identical
results. 
Here both, the experimental data and the calculations, are divided by the experimental values in order
 to magnify small differences. 
The error band for the calculations reflects a range of $\hbar\omega$  from 16 to 24~MeV, indicating that
both calculations are not converged at the $N_{\rm max}$=10 value used here. 
The calculations deviate on average from  the experimental values by about 5\% for both $^{16}$O and $^{12}$C. 
It is noteworthy to observe that the energy dependence of the calculated values of the total cross
sections slightly deviates from that given by experiment, being larger for $^{12}$C even in the energy
range between 100 and 200~MeV, which should be dominated by the leading order term in the multiple
scattering expansion.

\subsection{Observables for projectile energies smaller than 100~MeV}

Though the  leading order term in the multiple scattering expansion is expected to be valid
for  projectile kinetic energies larger than about 100~MeV, it is worthwhile to explore the behavior of
the leading order calculations at lower energies to study its energy dependence. 
For this study we use the Helium isotope chain together with $^{12}$C, and choose energies for which
experimental information is available.

\begin{figure}
\centering
\includegraphics[width=12cm]{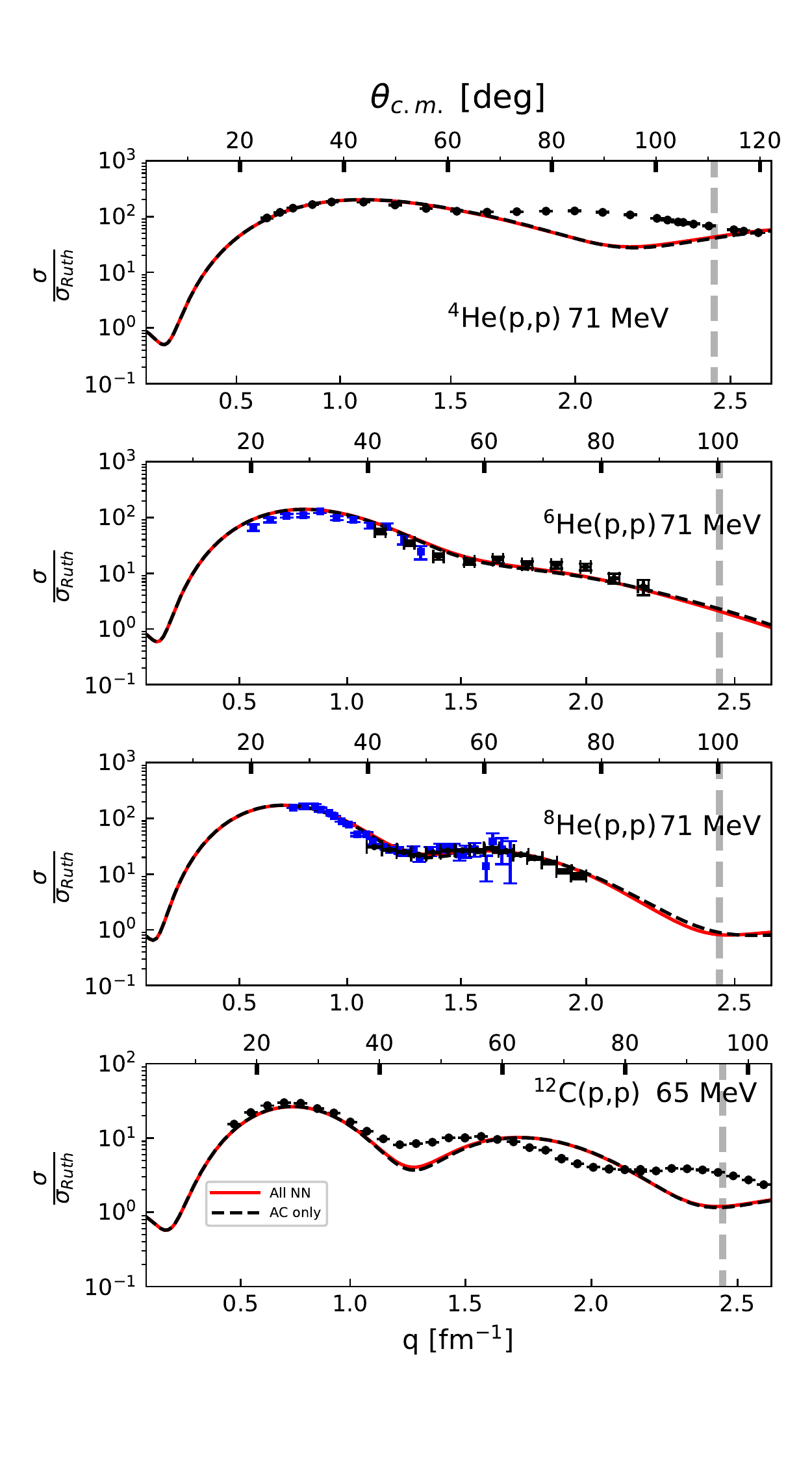}
\caption{The angular distribution of the differential cross section divided by the Rutherford cross
  section for elastic proton scattering from $^4$He, $^6$He, and $^8$He at 71~MeV laboratory kinetic energy and $^{12}$C at 65~MeV laboratory kinetic energy
as a function of the momentum transfer and the c.m. angle calculated with the NNLO$_{\rm opt}$
  chiral interaction~\protect\cite{Ekstrom13}. The meaning of the lines is the same as in
Fig.~\ref{fig3}. All calculations employ $\hbar \omega$=20 with $N_{\rm max}$=18 for $^4$He and $^6$He, $N_{\rm max}$=14 for $^8$He, and $N_{\rm max}$=10 for $^{12}$C. The square (blue) data points for $^6$He and $^8$He are taken from Ref.~\cite{Korsheninnikov:1997mm} while the circles (black) are taken from Refs.~\cite{Sakaguchi:2011rp,Sakaguchi:2013uut}.
}
\label{fig10}
\end{figure}

In Fig.~\ref{fig10}, the differential cross section divided by the Rutherford cross section for all Helium
isotopes is shown for projectile kinetic energy 71~MeV and for $^{12}$C 
at 65 MeV.  All line styles follow the same convention
given in Fig.~\ref{fig3}. 
We first notice  that the magnitude of the differential cross sections for the Helium isotopes 
 is predicted correctly for a momentum transfer up to about 2~fm$^{-1}$, slightly less for $^4$He. In the
case of $^{12}$C the magnitude of the differential cross section is still correctly predicted, but only
for momentum transfers up to 1~fm$^{-1}$. In addition, the first minimum for $^{12}$C is shifted to
a slightly higher momentum transfer with respect to the experimental values. In general, it is expected that for projectile energies smaller than 100~MeV corrections to the leading
order term~\cite{Chinn:1993zz,Chinn:1995qn} as well as higher order terms in the multiple scattering
expansions become important and are visible for higher momentum transfers. This can be seen in the
differential cross sections for $^4$He and $^{12}$C. The remarkable agreement of the leading order term
for the differential cross sections for $^6$He and $^8$He may be explained by the fact that those nuclei
are halo nuclei, and thus at those lower energies a large fraction of the scattering occurs from the
neutrons in the halo.

\begin{figure}
\centering
\includegraphics[width=12cm]{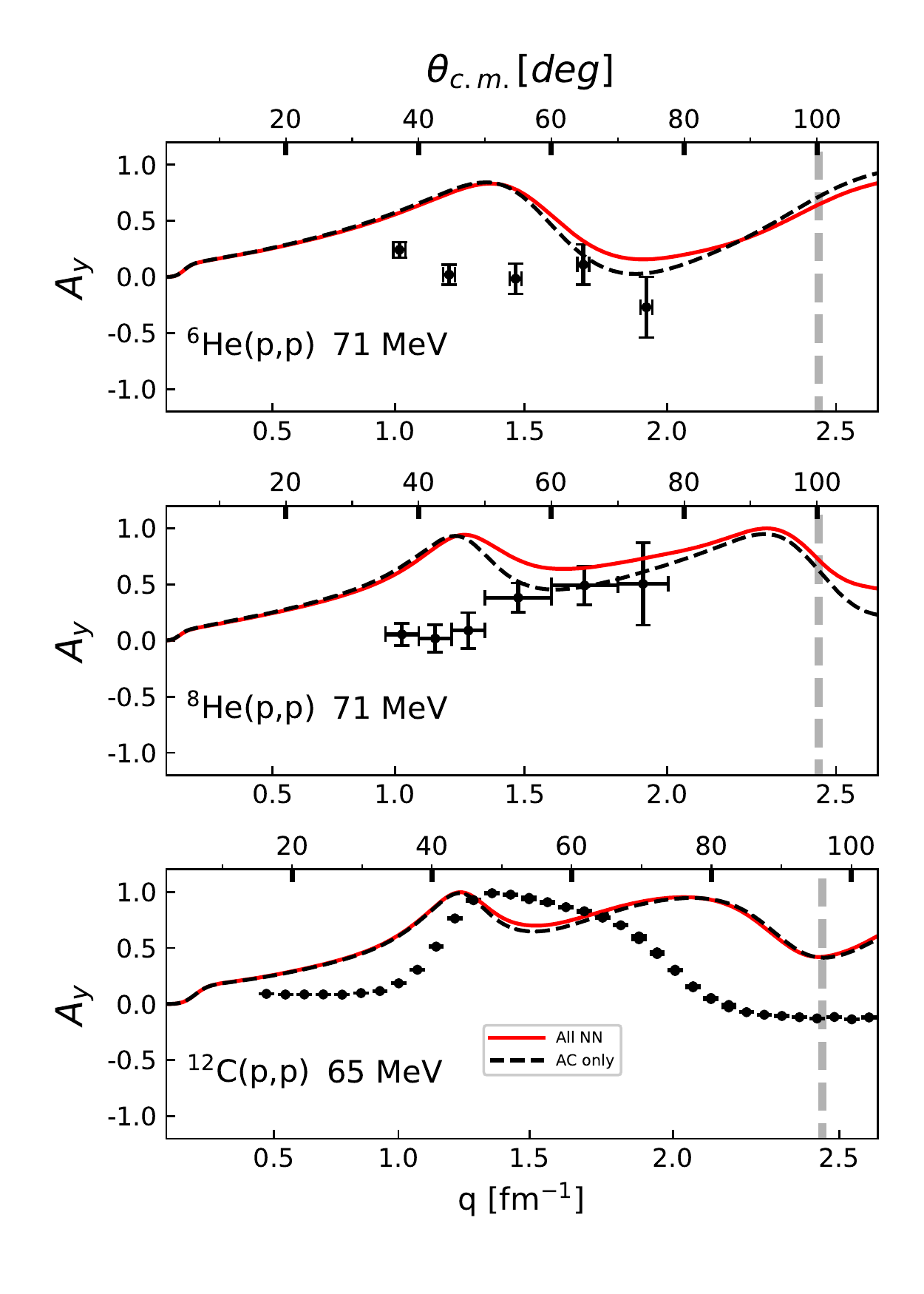}
\caption{Same as Fig.~\ref{fig10} but for the analyzing power. The data for $^6$He are taken from
Ref.~\cite{Uesaka:2010mm} and the data for $^8$He are from \cite{Sakaguchi:2013uut} \\
}
\label{fig11}
\end{figure}

While the differential cross sections exhibit no difference with respect to including or omitting the spin
of the struck nucleon in the nucleus, the analyzing powers do.
In Fig.~\ref{fig11}, the analyzing powers for the elastic scattering of protons off $^6$He, $^8$He, and
$^{12}$C are shown at the same energies. The  calculations confirm the  pattern already seen
for 100~MeV in Fig.~\ref{fig8}, where the nuclei with a larger $N/Z$ ratio are sensitive to treating the
spin of the struck nucleon correctly. Though the effects seen in $^6$He, $^8$He are most likely too small to be
experimentally verified, it is still important to point out that for nuclei with a larger $N/Z$ ratio the
spin of the struck nucleon should not be ignored.

\begin{figure}
\centering
\includegraphics[width=16cm]{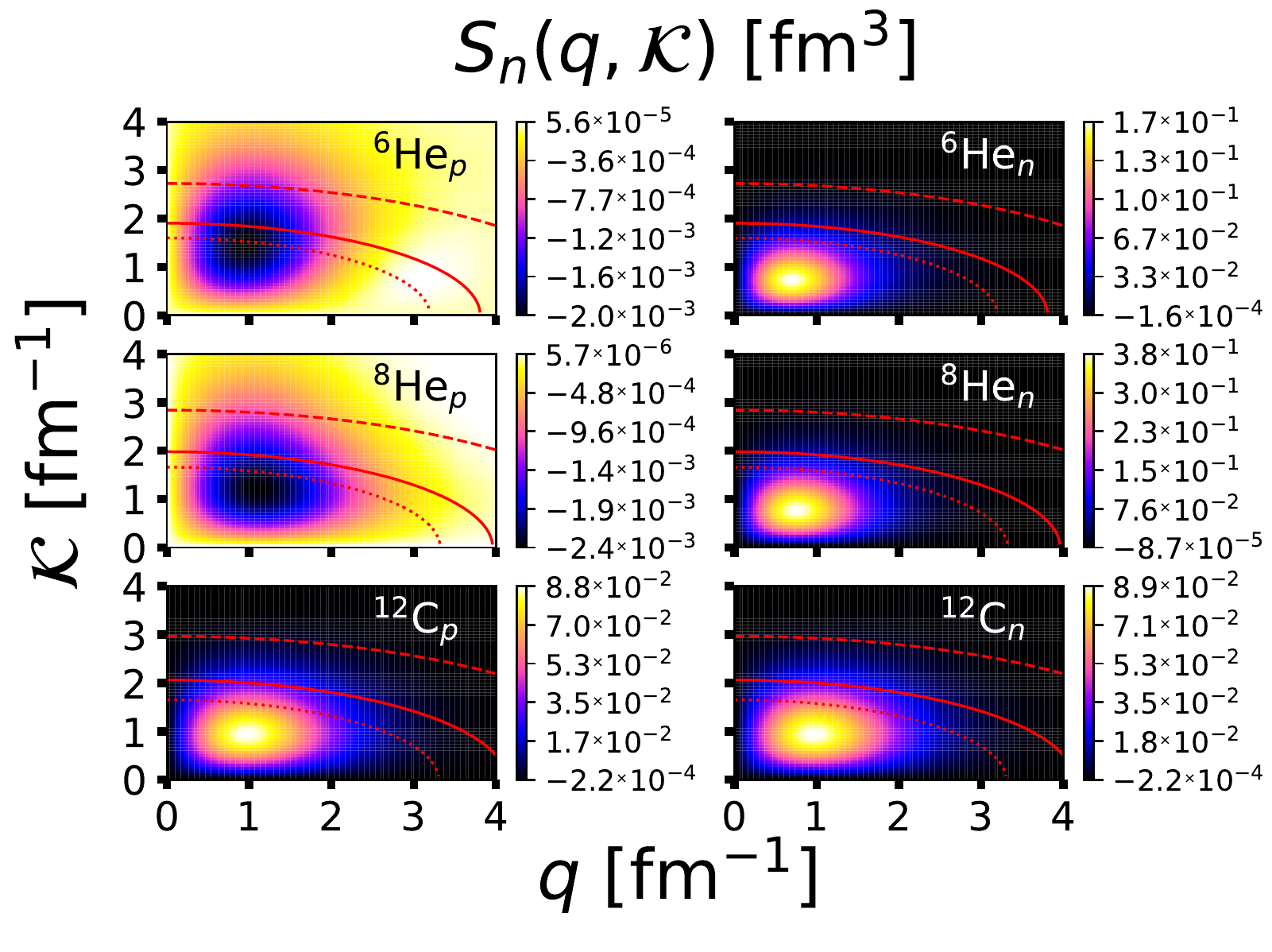}
\caption{The scalar function $S_n(\bm{q},\bm{\mathcal{K}})$ as function of the momentum transfers $q$
and ${\cal K}$ with $\cos \bm{q}\cdot \bm{\mathcal{K}} =0$  for $^6$He, $^8$He, and $^{12}$C. The
left column depicts $S_n$ calculated using the proton density, while the right column represents
$S_n$ derived from the neutron density. The dashed, solid, and dotted lines represent the on-shell
conditions for 200~MeV, 100~MeV, and 71~MeV (for $^{12}$C 65~MeV) respectively. 
}
\label{fig12}
\end{figure}

Finally, we show in Fig.~\ref{fig12} the intrinsic spin-projected momentum distribution Sn(q,K), as given in Eq.~(\ref{eq13}), as a function of the magnitudes of $\bm{q}$ and $\bm{\mathcal{K}}$ with the angle between the two vectors fixed at 90 degrees.  The scalar function $S_n(\bm{q},\bm{\mathcal{K}})$ is shown for the three nuclei $^6$He, $^8$He, and $^{12}$C with the proton spin-projected momentum distribution in the left panels and the neutron distribution in the right panels. These distributions show the effects of filling up the p-shell with either protons or neutrons and how that flips the sign from negative for the s-shell, as seen in the Helium isotopes, to positive for the p-shell.

The red bands shown on each plot represent three different on-shell momentum conditions, given through $q^2 +4\mathcal{K}^2 = 4k_0^2$, where $k_0$ is the momentum of the beam.  For $^6$He and $^8$He the dashed line is for 200 MeV, the solid line for 100 MeV, and the dotted line for 71 MeV while $^{12}$C follows the same scheme except that the dotted line is for 65 MeV. This shows clearly that the 100 and 71 MeV on-shell conditions are much closer to the peak of the spin-projected distributions while the 200 MeV line is significantly further out for both $^6$He and $^8$He.  This could explain the energy dependence seen in the previous scattering results and may indicate which nuclei will exhibit spin effects.

For $^6$He and $^8$He there is a disparity between the magnitude and shape of the neutron and proton momentum distributions, as one might expect. The spin-projected proton density of the alpha-core in $^6$He and $^8$He is significantly smaller than the spin-projected neutron  density in these nuclei, which is in agreement with the earlier observation that the spin-projected density does not play a role in $^4$He.  And not surprisingly, having 2 more neutrons than $^6$He, the spin-projected neutron distribution for $^8$He is twice as large, even though the spin-projected proton distribution is approximately the same in magnitude as in $^6$He.  For $^{12}$C on the other hand, the proton and neutron spin-projected densities are approximately the same, as one would expect for a $N$=$Z$ nucleus; furthermore, the shape of these nonlocal spin-projected distributions is somewhat similar to that of the corresponding neutron distributions of $^6$He and $^8$He.  However, even though $^{12}$C has 2 more neutrons than $^6$He, and the same number of neutrons as $^8$He, the magnitude of the spin-projected neutron distribution is half as large as that of $^6$He, and about a quarter of that of $^8$He.  This suggests that the detailed structure of the nucleus as well as the number of protons matters for the spin-projected momentum distributions, and could explain why ignoring the spin of the struck nucleon is a reasonable approximation in the case of $^{12}$C, but not for $^6$He an $^8$He. These results warrant further investigation into other nuclei and different interactions.

\section{Conclusions and Outlook}
\label{conclusions}

We calculated for the first time a complete leading order \textit{ab initio} effective potential for
nucleon-nucleus elastic scattering using the spectator expansion of multiple scattering theory. Complete
means here that we treat the $NN$ interaction in the reaction part of the calculation on the same footing as
in the structure part.  Taking the complete $NN$ interaction into account in the leading order term
implies that not only the spin of the projectile has to be considered but also the spin of the struck
target nucleon. In the context of full-folding effective interactions this has not been done according to
our knowledge, though the same idea was pioneered in the context of spin-spin terms in elastic scattering
from a target with non-zero spin~\cite{Cunningham:2011zz,Cunningham:2013lga}. 

In order to include the spin of the struck nucleon, we needed to explicitly include its spin operator
into the definition of the nonlocal density.  This is carried out by introducing a spherical
spin tensor of rank 1 into the definition of the density, 
allowing us to extract the usual scalar one-body density as well as a spin density.
To combine this with the structure of the $NN$ amplitudes given in the Wolfenstein representation, 
we needed to derive the projections of the spin
operator of the struck nucleon onto the three different linear independent momenta spanning the target
space. We found that for nucleon-nucleus scattering off 0$^+$ nuclei the projection of the spin along the normal of the plane spanned by the momentum transfer and the average momentum gives a non-vanishing result.
This  leads to an additional contribution of the Wolfenstein amplitude $C$ to the central part and
of Wolfenstein amplitude $M$ to the spin-orbit part of the effective potential.


We calculated proton elastic scattering observables for the closed-shell nuclei $^4$He and $^{16}$O at
multiple energies between 100 MeV and 200 MeV and compared to calculations in which the spin of the struck
nucleon is ignored. We find that the difference between the two is negligible. That confirms qualitative
arguments in traditional calculations that for closed-shell nuclei spin contributions most likely average
out.

Scattering observables for the  open-shell nuclei $^6$He, $^8$He, and $^{12}$C were also examined with
respect to their sensitivity 
to the spin of the struck nucleon. Each nucleus was examined between the energy range of 100 MeV to 200
MeV for proton elastic scattering. The results of this analysis show a trend of larger effects for lower
projectile energies as well as larger effects for nuclei that have a higher $N/Z$ ratio. These trends
however are not conclusive due to the small number of nuclei we examined. 

The differential cross sections at lower energies examined for the Helium isotopes along with $^{12}$C as
well as the analyzing powers for $^6$He, $^8$He, and $^{12}$C show similar energy dependence for the
contribution of the spin of the struck nucleon. A somewhat surprising result from  this study is that the 
differential cross sections for 
the halo nuclei $^6$He and $^8$He agree much better with experiment as one would expect for leading
order calculations. That could indicate that due to the loosely bound structure of a halo nucleus multiple
scattering effects appear at somewhat lower energies and higher momenta.  
However, to see if this is a general feature for
halo nuclei, one will need to study more cases. Analyzing powers usually give a more detailed view of the
effective interaction. Here it is quite obvious that a leading order calculation does not capture the measured structure at lower energies.

Summarizing, in this work we concentrated on pursuing the theoretical advancement of the description of
the leading order term in the multiple scattering series. Therefore, we only used a single $NN$ interaction,
the NNLO$_{\rm opt}$ interaction from Ref.~\cite{Ekstrom13}. In future work similar studies will
have to be carried out with different chiral interactions, as well as for e.g. another isotope chain to
further examine open-shell nuclei as the $N/Z$ ratio increases.



\appendix

\section{Spin-Projected Momentum Distribution}
\label{appendixA}
In this Appendix we show more details of the derivation of the function, $S_n\left(\bm{q}, \bm{\mathcal{K}}\right)$, that is related to the expectation value of $\sigma_i \cdot \hat{\bm{n}}$ in the ground state of the nucleus. The momentum vectors $\bm{q}$ and  $\bm{\mathcal{K}}$ are defined in Eq.~(\ref{eq6}). We start with the scalar function $S_n\left(\bm{p},\bm{p}' \right)$ defined in Eq.~(\ref{density}), with the one-body spherical spin tensor of rank $K_s = 0, 1$,
$\hat \sigma_{q_s}^{K_s}$, defined as
\begin{eqnarray}
\label{sigma}
K_s = 0~&:~~~~\left({\bm{\hat\sigma}}\right)^{0}_{0}  =& 1 \cr
 \cr
K_s = 1~&:~~~~\left({\bm{\hat\sigma}}\right)^{1}_{0}  =&  \bm{\sigma}_z \cr
&:~~\left({\bm{\hat\sigma}}\right)^{1}_{-1} =& \frac{1}{\sqrt{2}} \left( \bm{\sigma}_x - i\bm{\sigma}_y \right) \cr
&:~~~~\left({\bm{\hat\sigma}}\right)^{1}_{1}  =& -\frac{1}{\sqrt{2}} \left( \bm{\sigma}_x + i\bm{\sigma}_y \right)~.
\end{eqnarray}

Since for $K_s = 0$, Eq.~(\ref{density}) becomes the scalar density that we derived in previous work~\cite{Burrows:2017wqn}, we are going to show here derivations for $K_s = 1$. In this case the spin-projected momentum distribution will be:

\begin{eqnarray}
	S_n\left(\bm{p},\bm{p}' \right) = \sum_{q_s} \left\langle \Phi \left|
\sum_{i=1}^{A} \delta^3( \bm{p_i} - \bm{p}) \delta^3( \bm{p_i}' - \bm{p}')
{\hat \sigma}_{q_s}^{(i) K_s=1} \right| \Phi \right\rangle(-1)^{q_s} (\bm{\hat{n}^1_{t.i.}})_{-q_s}.
\end{eqnarray}

Following the procedure from~\cite{Burrows:2017wqn}, we expand the delta functions in terms of the spherical harmonics and couple them to bipolar harmonics, which we then couple to the spin tensor to get a total tensor of rank $K$ and  get the expression,

\begin{eqnarray}
\hspace*{-2cm} S_n (\bm{p},\bm{p'}) &=& \sum_{q_s}(-1)^{q_s}(\bm{\hat n_{t.i.}}^1)_{-q_s}
\sum_{\mu \mu'} \sum_{K_l=|\mu-\mu'|}^{\mu+\mu'} \sum_{k_l=-K_l}^{K_l} \mathcal{Y}_{K_lk_l}^{*\mu\mu'}(\hat{\bm{p}},\hat{\bm{p}'}) \sum_{Kk} \lla K_l k_l K_s q_s | K k \rra \cr
& & \times \lla  A \lambda J M \left| \sum_{i=1}^A \left[ \frac{\delta(p_i-p)}{p^2}
\frac{\delta(p_i'-p')}{p^{'2}} \mathcal{Y}_{K_lk_l}^{\mu\mu'}(\hat{\bm{p_i}},\hat{\bm{p_i}'}) 
\hat \sigma^{(i) 1}_{q_s} \right]^K_{k} \right| A \lambda J M \rra~.
\end{eqnarray}

We expand the tensor of rank $K$ in terms of single-particle matrix elements, 

\begin{eqnarray}
\label{A4}
\hspace*{-3cm} & &S_n(\bm{p},\bm{p'}) = \sum_{q_s}(-1)^{q_s}(\bm{\hat n_{t.i.}}^1)_{-q_s}
\sum_{\mu \mu'} \sum_{K_l=|\mu-\mu'|}^{\mu+\mu'} \sum_{k_l=-K_l}^{K_l} \mathcal{Y}_{K_lk_l}^{*\mu\mu'}(\hat{\bm{p}},\hat{\bm{p}}') \sum_{Kk} \lla K_l k_l K_s q_s | K k \rra (-1)^{J-M} \threej{J}{K}{J}{-M}{k}{M} \times \cr
& &\frac{1}{\hat{K}} \sum_{\alpha\beta} \lla \alpha \left|\left| \left[ \frac{\delta(p_1-p)}{p^2}
\frac{\delta(p_1'-p')}{p^{'2}} \mathcal{Y}_{K_l}^{\mu\mu'}(\hat{\bm{p_1}},\hat{\bm{p_1}'}) {\hat \sigma^1} \right]_{K} \right|\right| \beta \rra \lla A \lambda J \left|\left| (a^{\dagger}_{\alpha} \tilde{a}_{\beta})^{(K)} \right|\right| A \lambda J \rra~,
\end{eqnarray}
where $\alpha$ and $\beta$ represent the final and initial single particle states, respectively,  $(a^{\dagger}_{\alpha} \tilde{a}_{\beta})^{(K)}$ represent the single particle transition operator of rank $K$, and $\hat{K}$ is defined as $\hat{K} = \sqrt{2K+1}$.
After evaluating the reduced matrix elements we obtain:
\begin{eqnarray}
\label{A5}
S_n\left(\bm{p},\bm{p}' \right) &=& \sum_{q_s}(-1)^{q_s}(\bm{\hat n_{t.i.}}^1)_{-q_s} \sum_{nljn'l'j'} \sum_{K_l=|l-l'|}^{l+l'} \sum_{k_l=-K_l}^{K_l} \sum_{Kk} \lla K_l k_l K_s q_s | K k \rra \cr
& &(-1)^{J-M} \threej{J}{K}{J}{-M}{k}{M} \mathcal{Y}_{K k}^{*l l'}(\hat{\bm{p}},\hat{\bm{p}}') \cr
& & (-1)^{-l} \hat{j}\hat{j'} \hat{s} \hat{K_s} \hat{K_l} \ninej{l'}{l}{K_l}{s}{s}{K_s}{j'}{j}{K} R_{n'l'}(p') R_{nl}(p) \times \cr
& & (-i)^{l+l'} \lla A \lambda J \left|\left| (a^{\dagger}_{n'l'j'} \tilde{a}_{nlj})^{(K)} \right|\right| A \lambda J \rra~.
\end{eqnarray}

In order to obtain the translational invariant spin-projected momentum distribution, we are using the Talmi-Moshinsky transformation from the ($\bm{p}$, $\bm{p}'$) variable to the non-local variables ($q$, $\mathcal{K}$).
\begin{eqnarray}
	& &R_{n'l'}(p') R_{nl}(p) \mathcal{Y}_{K_l k_l}^{*l'l}(\hat{\bm{p}},\hat{\bm{p}}') = \cr
	& & \sum_{n_q,n_{\mathcal{K}},l_q,l_{\mathcal{K}}} \lla n_{\mathcal{K}} l_{\mathcal{K}}, n_q l_q : K_l | n' l', n l : K_l \rra_{d=1} R_{n_{\mathcal{K}} l_{\mathcal{K}}}(\mathcal{K}) R_{n_q l_q}(q) \mathcal{Y}_{K_l k_l}^{*l_{\mathcal{K}} l_q}(\hat{\bm{q}},\hat{\bm{\mathcal{K}}})~.
\end{eqnarray}

The intrinsic spin-projected momentum distribution, $S_n\left(\bm{q}, \bm{\mathcal{K}}\right)$ becomes:
\begin{eqnarray}
\label{SODM}
S_n\left(\bm{q}, \bm{\mathcal{K}}\right) &=&\sum_{q_s}(-1)^{q_s}(\bm{\hat n_{t.i.}}^1)_{-q_s}
\sum_{nljn'l'j'} \sum_{K_l=|l-l'|}^{l+l'} \sum_{k_l=-K_l}^{K_l} \sum_{Kk} \lla K_l k_l K_s q_s | K k \rra \cr
& &(-1)^{J-M} \threej{J}{K}{J}{-M}{k}{M} (-1)^{-l} \hat{j}\hat{j'} \hat{s} \hat{K_s} \hat{K_l} \ninej{l'}{l}{K_l}{s}{s}{K_s}{j'}{j}{K} (-i)^{l+l'} \cr
& & \sum_{n_q,n_{\mathcal{K}},l_q,l_{\mathcal{K}}} \lla n_{\mathcal{K}} l_{\mathcal{K}}, n_q l_q : K_l | n' l', n l : K_l \rra_{d=1} R_{n_{\mathcal{K}} l_{\mathcal{K}}}(\mathcal{K}) R_{n_q l_q}(q) \mathcal{Y}_{K_l k_l}^{*l_{\mathcal{K}} l_q}(\hat{\bm{q}},\hat{\bm{\mathcal{K}}}) \cr
& &\lla A \lambda J \left|\left| (a^{\dagger}_{n'l'j'} \tilde{a}_{nlj})^{(K)} \right|\right| A \lambda J \rra e^{\frac{1}{4A} b^2 q^2}~.
\end{eqnarray}
The term $e^{\frac{1}{4A} b^2 q^2}$ in this equation arises from the removal of the center-of-mass motion of the nucleus that follows the same procedure as in~\cite{Burrows:2017wqn}. The center-of-mass wavefunction is entirely in the $0s$ ground state of the nucleus. We obtained the same factor as in the scalar density.

It is important to notice that the first summation can be used to introduce a specific representation of the non-local momenta $\bm q$ and $\bm{\mathcal{K}}$ by using the definition of the independent vector $\hat{\bm n}$  from Eq.~(\ref{eq8}). By chosing the vector ${\bm q}$ in the $z$-direction and $\bm{\mathcal{K}}$ in the $x$-$z$-plane, the direction of $\hat{\bm n}$ is in the negative $y$-direction. Then using the spherical harmonics representation of the vector ($\hat{\bm n}_{\alpha} = \left| \hat{\bm{n}} \right| \sqrt{4\pi/3} Y^{1}_{\alpha}(\hat{\bm{n}})$), and taking into consideration the spatial configuration of $\hat{\bm n}$, the expression of  $S_n\left(\bm{q}, \bm{\mathcal{K}}\right)$  can be further simplified as given in Eq.~(\ref{eq13}).

\section{Frame Transformation and Projection}
\label{appendixB}

The transformations and projections between the different frames within the elastic scattering problem
are a complicated detail to accurately manage. Therefore,  we present here in detail  the derivation of
the angle  $\beta$ given in Sec.~\ref{formal}.

\begin{figure}
\centering
\includegraphics[width=100mm]{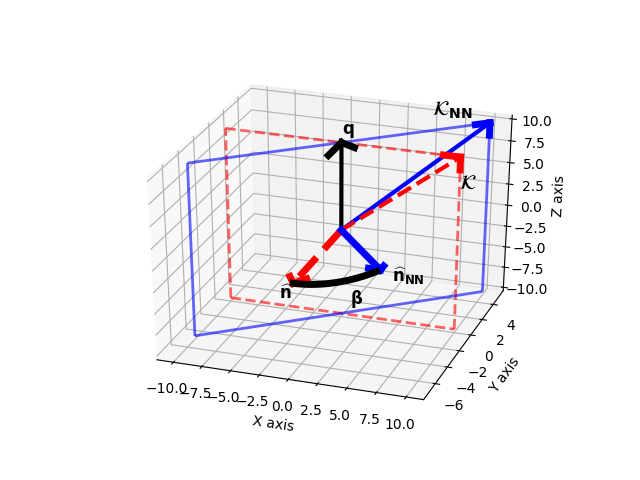}
\caption{The geometry of  the scattering planes of the $A$ frame and the \textit{NN} frame. The vectors
shown are the momentum transfer $\bm{q}$, the average momentum in the $A$ frame $\bm{\mathcal K}$, the
average momentum in the \textit{NN} frame $\bm{\mathcal K}_{NN}$, the normal vector ${\hat n}$ in the $A$
frame and the  normal vector ${\hat n}_{NN}$ in the $NN$ frame together with the angle $\beta$ between the two.
}
\label{Planes}
\end{figure}

We need to distinguish between three different frames, the nucleon-nucleus $NA$, the nucleon-nucleon $NN$,
and the target $A$ frame.
The  scattering problem is determined by two vectors, the momentum transfer $\bm{q}$ and the
average momentum $\bm{\mathcal{K}}$, leading to three variables: 
the magnitude of the momentum transfer, $\left| \bm{q} \right|$, the magnitude of the average momentum, $\left| \bm{\mathcal{K}} \right|$, and the angle in between them
$\theta_{q\mathcal{K}}$. The two  vectors form the scattering plane from which the unit vector
$\hat{\bm{n}}=\frac{\bm{\mathcal{K}} \times \bm{q}}{\left| \bm{\mathcal{K}} \times \bm{q}
\right|}$ is defined. The $NA$ frame and the $NN$ frame have each their own scattering plane with 
the angle between
scattering planes being defined as $\beta$. This geometry is shown in Fig.~\ref{Planes}.

Without loss of generality  we can choose the vector $\bm{q}$ to be parallel to the $z$-axis. We also note that $\bm{q}$ is invariant under frame
transformations.  Furthermore, we can choose the location of a specific scattering plane 
in the $x$-$z$-plane. Our choice here is the scattering plane in the $A$ frame, which in turn forces
the $\hat{\bm{n}}$ unit vector in the $A$ frame  to be along the negative $y$-axis.

The explicit definitions of the momentum transfer and average momentum in the $A$ frame are repeated here for convenience
\begin{eqnarray}
       \bm{q} &=& \bm{p'} - \bm{p} \cr
       \bm{\mathcal{K}} &=& \half(\bm{p'} + \bm{p}) \cr
       \widehat{\bm{n}} &=& \frac{\bm{p} \times \bm{p'}}{\left| \bm{p} \times \bm{p'} \right| } =
\frac{\bm{\mathcal{K}} \times \bm{q}}{\left| \bm{\mathcal{K}} \times \bm{q} \right| }~,
\end{eqnarray}
where $\bm{p}$ ($\bm{p'}$) are the the initial (final) momentum of the nucleon within the nucleus.
The functional form of the corresponding vector in the other frame is the same.

The relation between coordinates of the $NA$ frame and those of the $A$ are
\begin{eqnarray}
	\bm{q}_{NA} &=& \frac{A}{A-1} \left( \bm{p'} - \bm{p} \right) \cr
	\bm{\mathcal{K}}_{NA} &=& \frac{A}{A+1}\left[\left(\bm{k'} + \bm{k}\right) +
\frac{1}{2} \left(\bm{p'} + \bm{p}\right) \right]~,
\end{eqnarray}
where $\bm{k}$ ($\bm{k'}$) are the initial (final) momentum of the projectile in the $NA$ frame.

Lastly, the coordinates used in the $NN$ frame involve the projectile and struck nucleon, 
\begin{eqnarray}
	\bm{q}_{NN} &=& \left( \bm{k'}_{NN} - \bm{k}_{NN} \right) \cr
	\bm{\mathcal{K}}_{NN} &=& \half(\bm{k'}_{NN} + \bm{k}_{NN}) = \frac{1}{2} \left( \frac{A+1}{A} \bm{\mathcal{K}_{NA}} - \bm{\mathcal{K}} \right) \cr
	\widehat{\bm{n}}_{NN} &=& \frac{\bm{k}_{NN} \times \bm{k'}_{NN}}{\left| \bm{k}_{NN} \times \bm{k'}_{NN} \right| } = \frac{\bm{\mathcal{K}}_{NN} \times \bm{q}_{NN}}{\left| \bm{\mathcal{K}}_{NN} \times \bm{q}_{NN} \right|}~,
\end{eqnarray}
where $\bm{k}_{NN}$ ($\bm{k'}_{NN}$) are the initial (final) momentum of the projectile in the $NN$ frame
which differs from the momentum in the $NA$ frame. Using these definitions, the transformations between
frames and the projections of one frame onto another can be evaluated, see also~\cite{Orazbayev:2013}.

From Fig.~\ref{Planes}, one  recognizes that in order to determine the angle $\beta$,
the orientation of
each scattering plane in terms of its azimuthal $\phi$ coordinate must be known. Thus, we define
\begin{eqnarray}
\label{beta}
	\cos(\beta) = \cos(\phi_{NN} - \phi)~.
\end{eqnarray}
Since $\phi$ is a known quantity being integrated over,  only  $\phi_{NN}$ needs to be determined in order to obtain $\cos(\beta)$.

The average momentum of the $NN$ frame can be determined from the average momenta of the $A$ and $NA$
frames as
\begin{eqnarray}
	\bm{\mathcal{K}}_{NN} &=& \half \left( \frac{A+1}{A} \bm{\mathcal{K}_{NA}} - \bm{\mathcal{K}} \right)~.
\end{eqnarray}
Using this definition, we can write $\bm{\mathcal{K}}_{NN}$ in terms of its Cartesian coordinates and thus
obtain the angle $\phi_{NN}$ from the individual components,
\begin{eqnarray}
	\bm{\mathcal{K}}_{NN} &=& \half \left[ \frac{A+1}{A} \left( \begin{array}{c} \mathcal{K}_{NA}\sin(\theta_{NA}) \\ 0 \\ \mathcal{K}_{NA}\cos(\theta_{NA}) \end{array} \right) - \left( \begin{array}{c} \mathcal{K}\sin(\theta)\cos(\phi) \\ \mathcal{K}\sin(\theta)\sin(\phi) \\ \mathcal{K}\cos(\theta) \end{array} \right) \right] \cr
	&=& \half \left( \begin{array}{c} \frac{A+1}{A} \mathcal{K}_{NA}\sin(\theta_{NA}) - \mathcal{K}\sin(\theta)\cos(\phi) \\ -\mathcal{K}\sin(\theta)\sin(\phi) \\ \frac{A+1}{A} \mathcal{K}_{NA}\cos(\theta_{NA}) - \mathcal{K}\cos(\theta) \end{array} \right)~.
\end{eqnarray}
Using the definition $\tan(\phi_{NN})=\frac{y_{NN}}{x_{NN}}$ we obtain $\phi_{NN}$ as
\begin{eqnarray}
\label{phinn}
	&&\hspace*{-1.5cm}\phi_{NN} = \tan^{-1}\left( \frac{-\mathcal{K}\sin(\theta)\sin(\phi)}{\frac{A+1}{A} \mathcal{K}_{NA}\sin(\theta_{NA}) - \mathcal{K}\sin(\theta)\cos(\phi)} \right)~.
\end{eqnarray}
As long as the momentum vectors  in the $NA$  and $A$ frames are known, one can calculate $\phi_{NN}$. 
Thus Eq.~(\ref{phinn}) defines $\phi_{NN}$ which enters Eq.~(\ref{beta}).

\begin{acknowledgments}
This work was performed in part under the auspices of the U.~S. Department of Energy under contract
Nos. DE-FG02-93ER40756 and DE-SC0018223, and  
by the U.S. NSF (OIA-1738287 \& PHY-1913728).
The numerical computations benefited from computing resources provided
by Blue Waters (supported by the U.S. NSF, OCI-0725070 and ACI-1238993, and the state of Illinois), as well as the Louisiana Optical
Network Initiative and HPC resources provided by LSU ({\tt www.hpc.lsu.edu}), together with resources of the National Energy Research Scientific Computing Center, a DOE Office of Science User Facility supported by the Office of Science of the U.S. Department of Energy under contract No. DE-AC02-05CH11231. 

\end{acknowledgments}

\bibliography{denspot,clusterpot,ncsm}


\clearpage

\end{document}